\documentclass[11pt]{amsart}
\usepackage[a4paper,margin=2.8cm]{geometry}
\usepackage{graphicx}
\usepackage{subcaption}
\usepackage{multirow}
\usepackage{amsmath,amssymb,amsfonts}
\usepackage{amsthm}
\usepackage{mathrsfs}
\usepackage[title]{appendix}
\usepackage{xcolor}
\usepackage{textcomp}
\usepackage{booktabs}
\usepackage{braket}
\usepackage{algorithm}
\usepackage{algorithmicx}
\usepackage{algpseudocode}
\usepackage{listings}
\usepackage{comment}
\usepackage{bm}
\usepackage{tikz}
\usetikzlibrary{arrows.meta}
\usepackage[hidelinks]{hyperref}

\theoremstyle{plain}
\newtheorem{theorem}{Theorem}
\newtheorem{proposition}[theorem]{Proposition}
\newtheorem{lemma}[theorem]{Lemma}
\newtheorem{corollary}[theorem]{Corollary}

\theoremstyle{remark}
\newtheorem{example}{Example}
\newtheorem{remark}{Remark}
\theoremstyle{definition}
\newtheorem{definition}{Definition}

\newcommand{\Fq}{\mathbb{F}_q}
\newcommand{\Fp}{\mathbb F_q[x]/\langle f(x)\rangle}

\newcommand{\Prob}[1]{P\left( #1\right)}
\newcommand{\floor}[1]{\left\lfloor #1\right\rfloor}
\newcommand{\cev}[1]{\overleftarrow{#1}}

\begin{document}

\title[QSCs \& polynomial shifts]{Quantum synchronizable codes, polynomial and quasi-polynomial shifts}

\author{Edgar Mart\'inez-Moro}
\address{IMUVa-Mathematics Research Institute, University of Valladolid, Valladolid, Spain}
\email{edgar.martinez@uva.es}

\author{Miguel de los R\'ios de Antonio}
\address{IMUVa-Mathematics Research Institute, University of Valladolid, Valladolid, Spain}
\email{miguel.rios@uva.es}

\subjclass[2020]{81P73, 94B15, 94B05, 94B60}
\keywords{Polycyclic codes, Quasi-polycyclic codes, Quantum synchronizable codes}

\begin{abstract}
Quantum synchronizable codes protect quantum information against Pauli errors and against misalignments of the block boundaries. Almost all known constructions use cyclic codes, because the channel shifts the transmitted block cyclically. We study what happens when the algebraic shift of the code differs from the physical one. When one uses a twisted padding, constacyclic
codes synchronize at no cost, and they are the only polycyclic codes with this property for a linear padding. For the
other polycyclic codes, we measure the cost of each misaligned position, and we show that the
phase damage it causes can be corrected as erasures. Using annihilator duality, we extend the
construction to quasi-polycyclic codes, in which the cost is confined to a single block. We prove
that the usual quasi-cyclic families have a distance bound that does not grow with the length, and
show that quasi-polycyclic codes over repeated-root trinomials overcome it, giving synchronizable
codes with larger dimension and distance than the known synchronizable codes of the same length.
\end{abstract}

\maketitle

\section{Introduction}\label{sec:intro}

Block synchronization is a classical problem without a direct quantum analogue: reading a frame
marker is a measurement, and measuring disturbs the state. Fujiwara introduced \emph{quantum
synchronizable codes} (QSCs), which correct Pauli errors and, at the same time, misalignments of the
block boundaries~\cite{Fujiwara2,Fujiwara1}. The standard construction combines the CSS
framework~\cite{CSS1,CSS2} with a pair of nested cyclic codes $C_2^{\perp}\subseteq C_2\subseteq C_1$: the
misalignment is read in the class of $x^s$ modulo $g_2/g_1$ (the quotient of the generator polynomials of $C_2$ and $C_1$ respectively), and the order of $g_2/g_1$ governs the tolerated window.

Since then, QSCs have been obtained from many cyclic families: quadratic residue
codes~\cite{Fujiwara3}, repeated-root cyclic codes~\cite{Luo20181461,DinhLeNguyenTansuchat2021,DinhNguyenTansuchat2023}, codes arising from cyclotomic
constructions~\cite{Li201912,Shi2021727,Tiwari2023619,Kewat2025,Kewat2025b}, BCH codes with explicit
distance~\cite{Wang20251427,Liu20231751}, chain-containing codes~\cite{Xie2016414} and
others~\cite{Li2021342,Sun2023,Guenda2017393}; see~\cite{Shinde2026,Tansuwannont2026} for surveys and
recent variants. Negacyclic codes appear in~\cite{DinhNguyenYamaka2020,DinhLeNguyenTansuchat2021,LuoMaLin2019}, but the
synchronizable codes there are still obtained from cyclic codes; in~\cite{LuoMaLin2019}, for instance,
the $(u+v|u-v)$ construction turns a cyclic and a negacyclic code of length $n$ into a cyclic code of
length $2n$. As far as we know, the only construction
that leaves the cyclic setting is the quasi-cyclic family of QSCs from repeated-root
codes~\cite{DuMaLiu2023}, based on the quasi-cyclic codes of~\cite{Galindo2018}.

Cyclic codes are natural in this setting since, after the usual padding, the channel shifts the transmitted block
\emph{cyclically}, and cyclic codes are exactly those whose algebraic shift is the physical one. Richer
families, such as constacyclic and polycyclic codes~\cite{Permouth2009227}, codes associated with
trinomials~\cite{trinomials} and quasi-polycyclic codes~\cite{bag2026annihilatortwistedeuclideanduality},
often have better parameters, but their shift differs from the channel's one. The question we address in this work
is: \emph{how much does it cost to synchronize with a code whose shift is different from the channel's shift, and when
is it worth paying?}

Our contributions are the following. In Section~\ref{sec:channel},
we consider a two-sided model of misalignments, as the difference of two independent slip processes,
which gives a discrete Laplace law with closed formulas. In Section~\ref{sec:consta} we introduce
a \emph{twisted padding} that makes the physical shift act exactly as the constacyclic one; this means that
constacyclic codes synchronize at no cost as stated in Theorem~\ref{theo:consta1}, and binomials are shown to be
the only polynomials with this property for a linear padding; see Proposition~\ref{prop:onlybinomials}.
In the case of a general polycyclic code in an ambient space specified by the polynomial $f$, the \emph{twisted acyclicity}, denoted as $\rho(f)$, counts the virtual errors
created by each misaligned position. Thus,  we extend Fujiwara's construction to
all polycyclic codes in Theorem~\ref{theo:main1} and show that the phase damage caused by the
synchronization measurement occurs at known positions, so it is corrected as erasures. Then, 
by using annihilator duality~\cite{bag2026annihilatortwistedeuclideanduality}, we give a CSS construction
for quasi-polycyclic codes in physical coordinates (Theorem~\ref{theo:cssanh}) and QSCs in which only
one block pays the synchronization cost (Theorem~\ref{theo:main3}).
Finally,  a structural bound in
Proposition~\ref{prop:exp:cap} shows that the quasi-cyclic and quasi-constacyclic families
of~\cite{Galindo2018,DuMaLiu2023} have a distance that does not grow with the length, and we find
quasi-polycyclic codes over repeated-root trinomials that overcome it. For instance, we obtain synchronizable $[[98,54,7]]_7$ and $[[98,42,7]]_7$ codes, compared
with the $[[98,48,5]]_7$ and $[[98,28,5]]_7$ codes of~\cite{DuMaLiu2023}.

The paper is organized as follows. Section~\ref{polic} recalls polycyclic and quasi-polycyclic codes,
their dualities and the CSS construction. Section~\ref{sec:channel} introduces the channel model.
Section~\ref{sec:consta} presents the twisted padding and constacyclic QSCs.
Section~\ref{sec:displa} treats polycyclic QSCs and the cost of the misalignment.
Section~\ref{sec:general} extends the construction to quasi-polycyclic codes. Section~\ref{sec:experimental}
collects the codes found by computer search, compares them with cyclic, constacyclic, and quasi-cyclic
QSCs, and evaluates their probability of success under the channel model.

\section{Preliminaries}\label{polic}
 
 Let $\mathbb F_q$ be the finite field with $q$ elements. We will assume that the reader is familiar with what a linear code  $C$ over $\mathbb F_q$ is and its basic parameters. We will say that a linear code  $C$ is a $[n,k,d]$ code if its length is $n$, its dimension is $k$, and its minimum Hamming distance is $d$. We write $\|v\|$ for the Hamming weight of a vector $v$ and $d(u,v)=\|u-v\|$ for the Hamming distance.
In this paper, we will focus on cyclic codes and their generalizations including polycyclic codes and quasi-polycyclic codes. These codes are defined over an algebra $\mathcal A$ defined as
\begin{equation}\label{eq:algebra}
    \mathcal A=\mathbb F_q[x]/\langle f(x)\rangle,
\end{equation}where  $f$ is a monic polynomial. It is a well-known fact that the algebra $\mathcal A$ is semisimple if and only if the polynomial $f$ has simple roots, possibly in an extension of $\mathbb F_q$. 

\begin{definition} A \emph{polycyclic code} $C$ is an ideal in the algebra $\mathcal A$ given in Equation~\eqref{eq:algebra} and we will denote it as $C\triangleleft \mathcal A$. 
If $f(x)=x^n-1$, the code is said to be cyclic and constacyclic code if $f(x)=x^n-\lambda$ where $\lambda\in\mathbb F_q^{*}$. It is said to be a trinomial code if $f(x)=x^n-ax^i-b$ where $a,b\in\mathbb F_q^{*}$ and $1\leq i\leq n-1$.
\end{definition}

\begin{definition} We will  define a \emph{quasi-polycyclic code} $C$ of index $\ell\geq 1$ as an $\mathcal A$-submodule of $\mathcal A^\ell$, where  $\mathcal A$ is the algebra given in Equation~\eqref{eq:algebra}  and we will denote it as $C\triangleleft \mathcal A^\ell$. If $\mathcal A=\Fq[x]/\langle x^n-1\rangle$, such codes are called \textit{quasi-cyclic codes}.
\end{definition}

These two classes defined above are well-known in the literature and have been studied under different aspects over finite fields and finite rings; see, for example, \cite{guneri2021quasicyclic, bag2026annihilatortwistedeuclideanduality} and the references therein.
Given a linear code $C$ in $\mathbb F_q^N$, we will define the \emph{Euclidean dual} of $C$ and denote it by $C^\perp$ as the $\mathbb F_q$-linear code given by $C^\perp=\{u\in \mathbb F_q^N\mid u\cdot v=0\,\forall v\in C\}$. 
When we have the extra structure of $\mathcal A$  being an algebra, another duality can be stated based on the annihilator of an ideal; we will define this dual for a quasi-polycyclic code in $\mathcal A^\ell $ as $C^{\circ}=\{u\in\mathcal A^\ell \mid \varepsilon(u\cdot v)=0\ \text{for all } v\in C\}$, where $u\cdot v=\sum_{i=1}^{\ell}u_iv_i\in\mathcal A$ and $\varepsilon(a)$ is the constant term of $a$ (see~\cite[Definition~3.1]{bag2026annihilatortwistedeuclideanduality}). Since $C$ is an $\mathcal A$-module, this is the same as $u\cdot v=0$ for all $v\in C$. Both dualities are
connected; the Euclidean duality is just the annihilator duality twisted by the Gram matrix $G_{\mathcal A}$ of the \emph{annihilator form} on $\mathcal A^\ell$
\[
\langle {u},{v}\rangle_0= \Bigl(\sum_{i=1}^{\ell}(u_iv_i)(0)\Bigr) 
\quad u,v\in\mathcal A^\ell,
\]
where $(u_iv_i)(0)$ for $u_i,v_i\in \mathcal A$ is just the constant term of the reduced representative of $u_i(x)v_i(x)\mod f(x)$. The following result shows the relation: let $\Psi$ be the mapping from the representation of the code $C$ as a submodule of $\mathcal A^\ell$ to a $\mathbb F_q$-linear space in $\mathbb F_q^{n\ell}$.
\begin{theorem}[Proposition 4.2 in \cite{bag2026annihilatortwistedeuclideanduality}]\label{teo:twist}
For every quasi-polycyclic code $C$ in $\mathcal A^\ell$,
\[
\Psi(C^\circ)\cdot \mathcal G= 
\Psi(C)^{\perp},\qquad \text{and}\qquad
\bigl(\Psi(C)\cdot \mathcal G\bigr )^{\perp}=\Psi(C^{\circ}),
\]
where $\mathcal G=\mathrm{Id}_\ell\otimes G_\mathcal A$.
\end{theorem}

An \([[n, k]]\) \emph{quantum error-correcting code} (QECC)  encodes \(k\) logical qudits, i.e. the actual quantum data intended for computation, into a larger system of \(n\) highly entangled physical qudits (\(n > k\)), each of them a \(q\)-dimensional quantum system. Mathematically, the code defines a \(q^{k}\)-dimensional subspace (known as the code space) embedded within the larger \(q^{n}\)-dimensional Hilbert space of the physical system.
This multi-qudit overhead enables the non-destructive detection and correction of quantum errors, such as bit-flips (\(X\)), phase-flips (\(Z\)), and their combinations (\(Y\)).  While there are different ways of constructing QECCs, one of the most popular is the Calderbank–Shor–Steane (CSS) framework that was developed independently and simultaneously in two separate papers~\cite{CSS1, CSS2}. It allows us to obtain a quantum code from two classical linear codes fulfilling some restrictions.
\begin{theorem}[CSS Construction]\label{theo:css}
    Let $C_1$ and $C_2$ be two classical linear codes with parameters $[n,k_1,d_1]$ and $[n,k_2,d_2]$, respectively. If $C_2^\perp \subset C_1$, there exists an $[[n, k_1+k_2-n]]$ quantum code that corrects up to $\left\lfloor \frac{d_1-1}{2}\right\rfloor$ bit-flip errors and $\left\lfloor \frac{d_2-1}{2}\right\rfloor$ phase-flip errors.
\end{theorem}
For a thorough explanation of this construction, the reader can check it in~\cite{Nielse}. It could be extended to correct misalignment errors \cite{Fujiwara2} as it will be explained in the next section.

\section{Channel model and misalignment}\label{sec:channel}
As mentioned above, qudits can experience both bit-flip ($X$) and phase-flip ($Z$) errors, which are local, i.e., they affect only one qudit. The Pauli errors in $n$ qudits can be represented as a vector $(x|z)\in \mathbb F_q^{2n}$. In this representation, the coordinate $x_i$ is nonzero if the $i$-th qudit was affected by an $X$ error and $0$ if it was not. The same applies to the coordinates $z_i$ and the phase errors $Z$.
To model the probability of Pauli errors, we will rely on two assumptions: \textit{(i)} all the qudits share the probability of error ($p_x$ for bit-flips and $p_z$ for phase-flips), \textit{(ii)} the probabilities of the qudits suffering errors are independent of each other. Note that this second assumption fails when analyzing burst errors. 
The numbers $b$ of bit flips and $p$ of phase flips then follow binomial distributions:
$$P(b=j)=\binom{ n}{j}p_x^j(1-p_x)^{n-j},\qquad P(p=j)=\binom{ n}{j}p_z^j(1-p_z)^{n-j}.$$
The probabilities $p_x$ and $p_z$ depend on the hardware used, and each system can be described by different parameters, yet common threads appear across different types of technologies. One of these   is biased noise ($p_z\gg p_x$)\cite{evans2007error}. In many platforms, phase-flip errors are much more likely to occur than bit-flip errors, since dephasing does not require an exchange of energy with the environment~\cite{Puri2020}, and asymmetric quantum codes have been used to correct biased noise~\cite{Ioffe2007Asymmetric}. These codes have different minimum distances for $X$ and $Z$ errors and they can be built with Theorem~\ref{theo:css} and synchronized as in Theorem~\ref{theo:Fuji}.
Qudits can experience a broader class of errors when being sent across quantum channels, such as synchronization errors.  Suppose an array of $n$ qudits $(q_0,\ldots,q_{n-1})$ is sent through a quantum network and the receiver's and the sender's clocks are not synchronized, it is possible that the receiver does not note the arrival of the first qudit $q_0$. In this case, it receives a qudit array $(q_1,\ldots, q_{n-1},q_n)$, where $q_n$ is the first qudit of the next message. Such errors are called \emph{misalignments}. They can also be understood as the deletion of the first qudit coupled with an insertion of the last qudit.  Note that
misalignments are global errors, i.e. they do not affect one specific qudit but the whole qudit array. Therefore, they can be represented by a single integer $s\in\mathbb Z$. If $s<0$ the word is misaligned to the left, while if $s>0$ it is misaligned to the right. 

\begin{definition}
    A quantum code is called an \emph{$(a_l,a_r)$-synchronizable code} (QSC) if it can correct up to $a_l$ left misalignments and $a_r$ right misalignments.
\end{definition}

  To our knowledge,  no probabilistic model has been proposed in the literature  for this type of errors.  We model it as the result of two independent, memoryless slip processes: $R$ right slips and $L$ left slips, with $R\sim\operatorname{Geom}(1-p_r)$ and $L\sim\operatorname{Geom}(1-p_l)$ supported on the set of non-negative integers, that is 
$$\mathrm{P}(R=j)=(1-p_r)p_r^j,\quad \mathrm{P}(L=j)=(1-p_l)p_l^j.$$
and $s=R-L$. 

\begin{lemma}\label{lem:twosided}
The misalignment $s=R-L$ follows the \emph{two-sided (asymmetric discrete Laplace)
geometric law}
\[
\mathrm{P}({s=j})=c\cdot\begin{cases} p_r^{\,j}, & j\geq 0,\\ p_l^{\,|j|}, & j<0,\end{cases}
\qquad c=\frac{(1-p_r)(1-p_l)}{1-p_rp_l}.
\]
In particular, for integers $a_l,a_r\geq 0$,
\[
\mathrm{P}({-a_l\leq s\leq a_r})=1-\frac{(1-p_l)\,p_r^{a_r+1}+(1-p_r)\,p_l^{a_l+1}}{1-p_rp_l}.
\]
\end{lemma}
\begin{proof}
For $j\geq 0$, $\mathrm{P}({R-L=j})=\sum_{i\geq0}(1-p_r)p_r^{i+j}(1-p_l)p_l^{i}=c\,p_r^j$;
the case $j<0$ is symmetric. The second formula follows by summing the two geometric tails.
\end{proof}
\begin{corollary}
    In the symmetric case $p_r=p_l=p_s$, we obtain the discrete Laplace law
    $$\mathrm{P}({s=j})=\frac{1-p_s}{1+p_s}\,p_s^{|j|}$$ with $\mathbb E(s)=0$ and
\[
\Prob{s\neq 0}=\frac{2p_s}{1+p_s},\qquad
\Prob{|s|\leq K}=1-\frac{2\,p_s^{\lfloor K\rfloor+1}}{1+p_s}\ (K\ge 0),
\]
\[
\mathbb E|s|=\frac{2p_s}{1-p_s^2},\qquad
\operatorname{Var}|s|=\frac{2p_s(1+p_s^2)}{(1-p_s^2)^2},\qquad
\Prob{s\in[-a_l,a_r]}=1-\frac{p_s^{a_l+1}+p_s^{a_r+1}}{1+p_s}.
\]
\end{corollary}

\begin{definition}
Let $\mathbb{F}_q$ be a finite field and let $f(x) \in \mathbb{F}_q[x]$ be a polynomial such that $f(0) \neq 0$. The \emph{order}  of the polynomial $f(x)$, denoted by $\operatorname{ord}(f(x))$, is defined as the cardinality of the set of remainders of the powers of $x$ modulo $f(x)$, that is
\[
\operatorname{ord}(f(x)) = \bigl|\{ x^a \pmod{f(x)} \mid a \in \mathbb{N} \}\bigr|,
\]
or, equivalently, the least integer $e>0$ such that $f(x)\mid x^e-1$.
\end{definition}

\begin{theorem}[\cite{Fujiwara1}]\label{theo:Fuji}
    Let $C_1=\langle g_1(x)\rangle$ and $C_2=\langle g_2(x)\rangle$ be two cyclic codes with parameters $[n,k_1,d_1]$ and $[n,k_2,d_2]$, respectively. If $C_2^\perp\subset C_2\subset C_1$ and $0 <a_l+a_r < \operatorname{ord}\big (\frac{g_2(x)}{g_1(x)}\big )$, then there exists an $(a_l,a_r)$-$[[n+a_l+a_r, 2k_2-n]]$ quantum synchronizable code that corrects up to $\left\lfloor \frac{d_1-1}{2}\right\rfloor$ bit-flip errors and $\left\lfloor \frac{d_2-1}{2}\right\rfloor$ phase-flip errors.
\end{theorem}
The original construction in~\cite{Fujiwara2} requires $a_l+a_r<k_1-k_2$; the condition with the order of $g_2/g_1$, which allows at least as large windows, is due to~\cite{Fujiwara1}.

One of the main steps on Fujiwara's proof was the encoding. The first $a_l$ and the last $a_r$ qudits were repeated at the end and beginning of the word, respectively. This new state is denoted as $\ket{v}_{(a_l,a_r)}=(\underbrace{v_{n-a_r},\ldots,v_{n-1}}_{a_r},\underbrace{v_0,\ldots,v_{n-1}}_n,\underbrace{v_0,\ldots,v_{a_l-1}}_{a_l})$ where $v\in C_2/C_2^\perp$ is a coset representative, we will call these extra qudits \emph{padding qudits}. The qudit array is divided in three windows: left ($a_r$), central ($n$), and right ($a_l$). 
As a consequence, misalignments can be represented in the central window as a \textit{shift operator} $S$, which acts as a cyclic permutation on a vector given by
\begin{equation}
    S(v_0, v_1, \ldots, v_{n-1}) = (v_{n-1}, v_0, \ldots, v_{n-2}).
\end{equation}
This linear operator is equivalent to multiplication by $x$ on the ambient space of cyclic codes, $\Fq[x]/\langle x^n-1\rangle$.
\section{Constacyclic codes and twisted padding}\label{sec:consta}
Constacyclic codes are ideals in the algebra $\Fq[x]/\langle x^n-\lambda\rangle$ where $\lambda\neq 0$. In this ring, multiplication by $x$ is no longer associated with shift operator $S$, but with a \textit{constashift} operator $S_\lambda$ following
\begin{equation}
    S_\lambda(v_0, v_1, \ldots, v_{n-1}) = (\lambda v_{n-1}, v_0, \ldots, v_{n-2}).
\end{equation}

To adapt Fujiwara's construction, it is necessary to twist the padding by multiplying the padding  qudits by powers of $\lambda$.
\begin{definition}\label{def:twisted}
Let $v=(v_0,\ldots,v_{n-1})\in\Fq^{n}$ and extend it to a bi-infinite sequence $(u_j)_{j\in\mathbb Z}$ by
$$u_j=\lambda^{-\lfloor j/n\rfloor}\,v_{j\bmod n},\qquad j\in\mathbb Z,$$
i.e.\ $u_{j}=\lambda\,u_{j+n}$ for all $j$. For integers $a_l,a_r\geq 0$, the \emph{$\lambda$-twisted padding} of $v$ is
$$v^{\lambda}_{(a_l,a_r)}=(u_{-a_r},\ldots,u_{-1}\mid u_0,\ldots,u_{n-1}\mid u_n,\ldots,u_{n-1+a_l})\in\Fq^{\,n+a_l+a_r}.$$
When $a_l,a_r\leq n$ this is
$$v^{\lambda}_{(a_l,a_r)}=(\underbrace{\lambda v_{n-a_r},\ldots,\lambda v_{n-1}}_{a_r}\mid
\underbrace{v_0,\ldots,v_{n-1}}_{n}\mid
\underbrace{\lambda^{-1}v_0,\ldots,\lambda^{-1}v_{a_l-1}}_{a_l}).$$
\end{definition}

A misalignment of $s$ positions  means that the receiver reads the central window given by
$$W_s(v)=(u_{-s},u_{-s+1},\ldots,u_{n-1-s}),\qquad -a_l\leq s\leq a_r .$$

\begin{lemma}\label{lem:twisted}
For every $v\in\Fq^{n}$ and every $s\in[-a_l,a_r]$,
$W_s(v)=S_\lambda^{\,s}\,v$, 
 in other words, $W_s(v)$ is the coefficient vector of $x^{s}v(x)\bmod (x^n-\lambda)$.
\end{lemma}
\begin{proof}
Clearly $W_0(v)=v$. For any $s\in\mathbb Z$,
$W_{s-1}(v)=(u_{-s+1},\ldots,u_{n-s})$, hence
$$S_\lambda W_{s-1}(v)=(\lambda u_{n-s},u_{-s+1},\ldots,u_{n-1-s})=(u_{-s},u_{-s+1},\ldots,u_{n-1-s})=W_s(v),$$
since $u_{-s}=\lambda u_{n-s}$. If we apply induction on $s\geq 0$, we get $W_s=S_\lambda^s v$. In the case $s<0$, we
use the same identity in the backward direction, since $S_\lambda$ is invertible provided that $\lambda\neq0$.
\end{proof}

At the quantum level, the map $\ket{v}\mapsto\ket{v^{\lambda}_{(a_l,a_r)}}$ is an isometry, since
$v\mapsto v^\lambda_{(a_l,a_r)}$ is an $\Fq$-linear map and it is injective. Moreover, it can be implemented with the
same controlled-addition gates as in~\cite{Fujiwara1}, followed by the multiplication gates
$\ket{a}\mapsto\ket{\lambda^{\pm1}a}$ on the padding qudits.

\begin{theorem}\label{theo:consta1}
 Let $C_1=\langle g_1(x)\rangle$ and $C_2=\langle g_2(x)\rangle$ be two $\lambda$-constacyclic codes with parameters $[n,k_1,d_1]$ and $[n,k_2,d_2]$, respectively. If $C_2^\perp\subset C_2\subset C_1$ and $0 <a_l+a_r < \operatorname{ord}(\frac{g_2}{g_1})$, then there exists an $(a_l,a_r)$-$[[n+a_l+a_r, 2k_2-n]]$ quantum synchronizable code that corrects up to $\left\lfloor \frac{d_1-1}{2}\right\rfloor$ bit-flip errors and $\left\lfloor \frac{d_2-1}{2}\right\rfloor$ phase-flip errors.
\end{theorem}

\begin{proof} As in~\cite{Fujiwara1}, the proof is divided into two main phases: encoding (A) and decoding (B), each with three subphases. We will explain all the phases for completeness. Throughout the proof, we will stick with a single ket $\ket c$ with $c\in C_2$; the argument extends by linearity to any linear combinations of states. 
\begin{itemize}
    \item[A]  Encoding.
    \begin{enumerate}
        \item 
    The CSS construction is used on $C_2$ to encode the information. The result is a sum of kets $\sum_i \alpha_i\ket {v_i+C_2^\perp}$ where the $v_i$ run over a set of representatives of the cosets in $C_2/C_2^\perp$.   
    \item To encode the alignment information, we add the generator $g_1$ to the state; therefore, the resulting ket is $\ket{c+g_1}$ with $c+g_1\in C_1.$   
    \item Twisted padding is applied to the message resulting   $\ket{c+g_1}_{(a_l,a_r)}^\lambda$  which is   sent across the quantum channel. 
    \end{enumerate}
    \item[B]  Decoding.
    \begin{enumerate}
        \item  Suppose that the message is received with phase flips given by a vector $e_Z$ of weight $p$, bit flips given by a vector $e_X$ of weight $b$, and a misalignment $s$. If $s\in [-a_l,a_r]$, by Lemma~\ref{lem:twisted} the   central window is
    $$\omega^{\operatorname{tr}(\langle e_Z,c+g_1\rangle)}S_\lambda^s\ket{c+g_1+e_X}.$$
    Now,  the parity check matrix $H_1$ of the code $C_1$ is applied
    $H_1S_\lambda^s(c+g_1+e_X)=H_1S_\lambda^s(c+g_1)+H_1S_\lambda^s e_X=H_1S_\lambda^s e_X$, because $C_1$ is $\lambda$-constacyclic, and
    the bit flips $S_\lambda^s e_X$, which have the same weight $b$ as $e_X$, can be corrected as long as $b\leq \left\lfloor \frac{d_1-1}{2}\right\rfloor$.
    \item   As $c\in  C_2$ it can be expressed as $\mu g_2$ for some polynomial class $\mu $. Thus, the state can be expressed as
    $$S_\lambda^s(c+g_1)=x^s(\mu  g_2+g_1)=x^s\left(\mu  \frac{g_2}{g_1}g_1+g_1\right).$$
    The syndrome of $C_2$ is now measured on the central window. Since $x^s\mu g_2\in C_2$, it equals the syndrome of $x^sg_1$ for every term of the superposition, so the measurement does not disturb the state, and it determines $x^sg_1\bmod g_2$, that is, $g_1\cdot\bigl(x^s\bmod \frac{g_2}{g_1}\bigr)$. Since $s\in[-a_l,a_r]$ and $a_l+a_r<\text{ord}\left(\frac{g_2}{g_1}\right)$, this determines $s$ uniquely. The alignment is then corrected by shifting the frame of the receiver by $s$ positions on the incoming stream of qudits, as in Section III.D in~\cite{Fujiwara2}; this may require gathering $|s|$ more qudits, and on the central window it acts as $S_\lambda^{-s}$. 
    \item Finally,  the padding qudits are deallocated since they are no longer needed. The errors act on the $n+a_l+a_r$ qudits of the transmitted block; when the padding is removed, $X$ errors on padding qudits stay there and are discarded, while a $Z$ error on a padding qudit is moved to the qudit it copies, so the number of qudits with phase flips in the central window does not increase. The parity check matrix $H_2$ of the code $C_2$ and the phase parity check matrix of $C_2^\perp$ given by $H_2^\perp$ are applied to the new central window. 
    \end{enumerate}
\end{itemize}
\end{proof}

Applying Theorem~\ref{theo:css} to the pair $C_2^\perp\subset C_3$, as is done for asymmetric
quantum codes~\cite{Ioffe2007Asymmetric,SKR2009}, instead of the self-orthogonal condition
$C_2^\perp\subset C_2$ in step (A1) of the proof of Theorem~\ref{theo:consta1} yields a more
general result. Note that the new code included $C_2$ needs only to be an $\mathbb F_q$-linear code; there is no need for it to be constacyclic.

\begin{theorem} \label{theo:consta2}
    Let $C_1=\langle g_1(x)\rangle$ and $C_3=\langle g_3(x)\rangle$ be two $\lambda$-constacyclic codes with parameters $[n,k_1,d_1]$ and $[n,k_3,d_3]$, respectively and $C_2$ be a $\Fq$-linear code with parameters $[n,k_2,d_2]$.  If $C_2^\perp\subset C_3\subset C_1$ and $a_l+a_r <\text{ord}\left( \frac{g_3}{g_1}\right)$, then there exists an $(a_l,a_r)$-$[[n+a_l+a_r, k_3+k_2-n]]$ quantum synchronizable code that corrects up to $\left\lfloor \frac{d_1-1}{2}\right\rfloor$ bit-flip errors and $\left\lfloor \frac{d_2-1}{2}\right\rfloor$ phase-flip errors.
\end{theorem}
\begin{proof}
    Only Step (A1) in~Theorem~\ref{theo:consta1} is modified, as follows. 
    The information is encoded with the CSS construction in Theorem~\ref{theo:css}
    applied to the pair $C_2^\perp\subset C_3$, as for asymmetric quantum
    codes~\cite{Ioffe2007Asymmetric,SKR2009}. Thus, the encoded state is a superposition
    $\sum_i \alpha_i\ket{v_i+C_2^\perp}$, where the $v_i$ run over a set of representatives of the
    cosets in $C_3/C_2^\perp$.
\end{proof}
In the case $\lambda=1$ and $C_3=C_2$ we recover Theorem~\ref{theo:Fuji}.

\begin{remark}\label{rem:dualconsta}
The Euclidean dual of a $\lambda$-constacyclic code is $\lambda^{-1}$-constacyclic, hence 
Euclidean dual-containing $\lambda$-constacyclic codes with $\lambda\neq1$ occur for
$\lambda=-1$. In other words, for the self-orthogonal setting of Theorem~\ref{theo:consta1}, the relevant case is
the negacyclic one.  General $\lambda$ can be used in the asymmetric setting of Theorem~\ref{theo:consta2}.
 Also, 
if $\lambda=\mu^{n}$ for some $\mu\in\Fq^{*}$, the map $a(x)\mapsto a(\mu^{-1}x)$ sends
$\Fq[x]/\langle x^n-1\rangle$ onto $\Fq[x]/\langle x^n-\lambda\rangle$, hence $\lambda$-constacyclic codes are
monomially equivalent to cyclic codes and nothing new is obtained. New codes can only appear
when $\lambda$ is not an $n$-th power in $\Fq^{*}$, which requires $\gcd(n,q-1)>1$.
\end{remark}

\begin{example}\label{ex:consta} In
Table~\ref{tab:consta} we compare, for a small range of lengths, the largest minimum distance among all
cyclic codes with the largest minimum distance among all $\lambda$-constacyclic codes of the
same length and dimension. Both were obtained by exhaustive search over all divisors of
$x^n-1$ and $x^n-\lambda$. In all these cases $\lambda$ is not an $n$-th power (Remark~\ref{rem:dualconsta}),
and the constacyclic codes are strictly better.

\end{example}

\begin{table}[h]
\centering
\begin{tabular}{cccc}
\toprule
$q$ & $\lambda$ & best cyclic & best $\lambda$-constacyclic\\
\midrule
$3$ & $-1$ & $[4,2,2]$ & $[4,2,3]$ \\
$3$ & $-1$ & $[10,4,4]$, $[10,6,2]$ & $[10,4,6]$, $[10,6,4]$ \\
$3$ & $-1$ & $[14,6,4]$, $[14,8,2]$ & $[14,6,6]$, $[14,8,5]$ \\
$5$ & $2$  & $[6,2,4]$, $[6,4,2]$ & $[6,2,5]$, $[6,4,3]$ \\
\bottomrule
\end{tabular}
\caption{Cyclic versus constacyclic codes.}
\label{tab:consta}
\end{table}

\section{Polycyclic codes and QSCs}\label{sec:displa}
Polycyclic codes are  ideals in the algebra $\Fq[x]/\langle f(x)\rangle$ where $f(x)=x^n+f_{n-1}x^{n-1}+\ldots+f_0$ is a monic polynomial with $f_0\neq 0$. In this ring, multiplication by $x$ is no longer associated with shift operator $S$, but with a \textit{polyshift} operator $S_f$ following
\begin{equation}
    S_f(v_0, v_1, \ldots, v_{n-1}) = (0, v_0, \ldots, v_{n-2})-\vec fv_{n-1},
\end{equation}
where $\vec f=(f_0,\ldots,f_{n-1}).$  As $f_0\neq 0$ this operator is invertible with
\begin{equation}
    S_f^{-1}(v_0, v_1, \ldots, v_{n-1}) = ( v_1, \ldots, v_{n-1},0)-\cev{ f}\frac{v_{0}}{f_0},
\end{equation}
where $\cev{f}=(f_1,\ldots,f_{n-1},1)$.
Most polyshift operators are incompatible with twisted padding and,   as we show in the following result, only constashifts are fully compatible with the physical shift on a qudit array.

\begin{proposition}\label{prop:onlybinomials}
Let $f=x^n+f_{n-1}x^{n-1}+\cdots+f_0\in\Fq[x]$ with $f_0\neq0$. If there is a
linear map $v\mapsto (w_{-1},v_0,\ldots,v_{n-1})$ such that the window shifted by one position
satisfies $(w_{-1},v_0,\ldots,v_{n-2})=S_fv$ for every $v\in\Fq^{n}$, then $f=x^n-\lambda$ for some
$\lambda\in\Fq^{*}$.
\end{proposition}
\begin{proof}
Let $x^n\equiv-(f_0+\cdots+f_{n-1}x^{n-1})$, then the entries of $S_fv$ are
$v_{i-1}-f_iv_{n-1}$ which must equal $v_{i-1}$ for every $v$ to fulfill the condition. Taking $v=e_{n-1}$, the vector with one in the $(n-1)$th position and zero elsewhere, we get 
$f_1=\cdots=f_{n-1}=0$, so $f=x^n+f_0$ and the result follows.
\end{proof}
We now measure how far a polyshift is from the physical shift; this is done by the twisted acyclicity of $f$. This magnitude represents how far the algebra $\Fq[x]/\langle f(x)\rangle$ is from being constacyclic.

\begin{definition}\label{def:twacyc}
The \emph{twisted acyclicity} of $f(x)$ is
$$\rho(f)=\min_{\lambda\in\Fq^{*}}\bigl\|\vec f+\lambda e_0\bigr\|
=\|\vec f\|-1 ,$$
where $e_0=(1,0,\ldots,0)$, so that $\vec f+\lambda e_0$ is the coefficient vector of $f-(x^n-\lambda)$. The minimum is attained exactly at $\lambda=-f_0$. We write
$\tilde f_{\lambda}=f-(x^n-\lambda)=f_1x+\cdots+f_{n-1}x^{n-1}$ for this choice of $\lambda$.
\end{definition}

Synchronization can still be attained with polycyclic codes, not strictly constacyclic codes, at the cost of $\rho(f)$  virtual errors for each misalignment. Before proving it we need a few technical lemmas.
\begin{lemma}\label{lemma:Stwisted}
Let $f\in\Fq[x]$ be monic of degree $n$ with $f_0\neq0$, and let $\lambda=-f_0$. For every
$v\in\Fq^{n}$ and every $s\in\mathbb Z$,
$$d\bigl(S_\lambda^{s}v,\,S_f^{s}v\bigr)\leq |s|\,\rho(f).$$
\end{lemma}
\begin{proof}
With the convention $x^n\equiv-(f_0+f_1x+\cdots+f_{n-1}x^{n-1})$,
$S_fv=(0,v_0,\ldots,v_{n-2})-v_{n-1}(f_0,f_1,\ldots,f_{n-1})$ and
$S_\lambda v=(-f_0v_{n-1},v_0,\ldots,v_{n-2})$, hence 
 $S_\lambda v-S_fv=v_{n-1}(0,f_1,\ldots,f_{n-1})$, which has weight at most $\rho(f)$.

For proving the inverse part, solving $S_fv=w$ gives $v_{n-1}=-f_0^{-1}w_0$ and
$v_{i-1}=w_i-f_if_0^{-1}w_0$ for $1\leq i\leq n-1$. On the other hand
$S_\lambda^{-1}w=(w_1,\ldots,w_{n-1},\lambda^{-1}w_0)$ with $\lambda^{-1}=-f_0^{-1}$. Therefore
$S_\lambda^{-1}w-S_f^{-1}w=f_0^{-1}w_0\,(f_1,\ldots,f_{n-1},0)$, again of weight at most $\rho(f)$.
For $s\geq2$, since $S_\lambda$ is monomial and hence a Hamming isometry,
$$d(S_\lambda^{s}v,S_f^{s}v)\leq d\bigl(S_\lambda S_\lambda^{s-1}v,\,S_\lambda S_f^{s-1}v\bigr)
+d\bigl(S_\lambda S_f^{s-1}v,\,S_fS_f^{s-1}v\bigr)\leq (s-1)\rho(f)+\rho(f),$$
by induction. The case where $s\leq-2$ can be proven in a similar way using  $S_\lambda^{-1}$ and $S_f^{-1}$.
\end{proof}

We are interested in codes with low twisted acyclicity. More concretely, we will use  constacyclic (binomial) and trinomial codes:
\begin{enumerate}
\item $\rho(f)=0$ if and only if $f=x^n-\lambda$, i.e.\ the code is constacyclic.
\item $\rho(f)=1$ if and only if $f=x^n+ax^{i}+f_0$ with $a,f_0\neq0$ and $1\leq i\leq n-1$,
i.e.\ $f$ is a trinomial.
\end{enumerate}
We need two further lemmas, the first describes the projection operator induced by syndrome measurements, while the second establishes error-and-erasure correction bounds for CSS codes. 
\begin{lemma}\label{lem:meas}
Let $\omega=e^{2\pi i/p}$, $\operatorname{tr}:\Fq\to\mathbb F_p$ the absolute trace and
$Z(c)\ket{y}=\omega^{\operatorname{tr}(c\cdot y)}\ket{y}$ for $c,y\in\Fq^n$. Let
$J\subseteq\{0,\ldots,n-1\}$, let $H_J:\Fq^{J}\to\Fq^{m}$ be linear and $\sigma\in\Fq^m$,then the diagonal
projector onto $\{y:H_J(y_J)=\sigma\}$ is
$$Q=\sum_{y: H_J(y_J)=\sigma}\ket{y}\!\bra{y}
=\frac1{q^{m}}\sum_{\beta\in\Fq^{m}}\omega^{-\operatorname{tr}(\beta\cdot\sigma)}\,Z\bigl(\iota_JH_J^{\!\top}\beta\bigr),$$
where $\iota_J$ extends a vector of $\Fq^J$ by zero. In particular, $Q$ is a linear combination of
phase-flip operators supported on $J$. 
\end{lemma}
\begin{proof}
The additive characters of $\Fq$ are $x\mapsto\omega^{\operatorname{tr}(\beta x)}$, $\beta\in\Fq$, and
$\sum_{\beta\in\Fq^m}\omega^{\operatorname{tr}(\beta\cdot u)}=q^m$ if $u=0$ and $0$ otherwise
\cite[Chapter~5]{LidlNiederreiter}. Apply this to $u=H_J(y_J)-\sigma$ and use
$\beta\cdot H_J(y_J)=(\iota_JH_J^{\!\top}\beta)\cdot y$.
\end{proof}

\begin{lemma}\label{lem:erasures}
Let $\mathcal C$ be the CSS code of $C_2^\perp\subseteq C_3$, so that bit flips are governed by
$d_x=d(C_3)$ and phase flips by $d_z=d(C_2)$. Let $J_x,J_z\subseteq\{0,\ldots,n-1\}$ be known sets. Then
every operator in the span of
$$\{X(a'+a'')Z(c'+c''):\ \|a'\|\le b,\ \operatorname{supp}(a'')\subseteq J_x,\
\|c'\|\le p,\ \operatorname{supp}(c'')\subseteq J_z\}$$
is correctable, i.e.\ $b$ bit flips and $p$ phase flips at unknown positions together with
arbitrary $X$-errors on $J_x$ and arbitrary $Z$-errors on $J_z$, provided
$$2b+|J_x|\le d_x-1\qquad\text{and}\qquad 2p+|J_z|\le d_z-1 .$$
\end{lemma}
\begin{proof}
Two such operators differ by $X(u)Z(v)$ up to phase with $\|u\|\le2b+|J_x|<d_x$ and
$\|v\|\le2p+|J_z|<d_z$. If $u\neq0$ then $u\notin C_3$, so $X(u)Z(v)$ maps the code to its orthogonal
complement. If $u=0\neq v$ then $v\notin C_2=(C_2^\perp)^\perp$, so $y\mapsto\omega^{\operatorname{tr}(v\cdot y)}$
is a non-trivial character on $C_2^\perp$ and $\Pi Z(v)\Pi=0$. Hence the Knill--Laflamme conditions hold,
and the recovery extends to the span by linearity; see~\cite{erasures}.
\end{proof}

We are now in a position to state the result for polycyclic codes with the penalization arising from $\rho$.
\begin{theorem}\label{theo:main1}
    Let $C_2=\langle g_2(x)\rangle$ and $C_1=\langle g_1(x)\rangle$ be two polycyclic codes over the algebra $\Fq[x]/\langle f(x)\rangle$ with $f_0\neq 0$. Let  $[n,k_1,d_1]$ and $[n,k_2,d_2]$ be the parameters of $C_1$ and $C_2$ respectively. If $C_2^\perp\subset C_2\subset C_1$ and $a_l+a_r <\text{ord}\left( \frac{g_2}{g_1}\right)$, then there exists an $(a_l,a_r)$-$[[n+a_l+a_r, 2k_2-n]]$ quantum synchronizable code that corrects up to $b$ bit-flips, $p$ phase-flips and $s$ misalignments if
    $$
        b+|s|\rho(f)\leq \left\lfloor \frac{d_1-1}{2}\right\rfloor,\quad
        2p+|s|\leq d_2-1,\quad
        s\in [-a_l,a_r].
    $$
 \end{theorem}
    \begin{proof}
The proof   follows the same schema as the one for Theorem~\ref{theo:consta1}.
\begin{itemize}
\item[A] Encoding.
\begin{enumerate}
    \item  The CSS construction is used on $C_2$ to encode the information, hence we get $\sum_i \alpha_i\ket {v_i+C_2^\perp}$ where the $v_i$ run over a set of representatives of the cosets in $C_2/C_2^\perp$. Unlike in Theorem~\ref{theo:consta1}, we cannot argue with a single ket: the syndrome measured in step B1 depends on the state, so we keep track of the whole superposition and write $\ket{c}$ for any of its terms.
    \item   To encode the alignment information, the generator $g_1$ is added  as $\ket{c+g_1}$ with $c+g_1\in C_1.$  
    \item  Twisted padding is applied to the message; hence, we obtain $\ket{c+g_1}_{(a_l,a_r)}^\lambda$  with $\lambda=-f_0$, which is sent across the quantum channel. 
    \end{enumerate}
  \item[B] Decoding.
\begin{enumerate}
    \item   
    Suppose that the message is received with phase flips given by a vector $e_Z$ of weight $p$, bit flips given by a vector $e_X$ of weight $b$, and a misalignment $s$. If $s\in [-a_l,a_r]$, by Lemma~\ref{lem:twisted} the resulting central window is 
    $$\omega^{\operatorname{tr}(\langle e_Z,c+g_1\rangle)}S_\lambda^s\ket{c+g_1+e_X}.$$
    Now, applying Lemma~\ref{lemma:Stwisted}  $d\bigl(S_\lambda^{s}v,\,S_f^{s}v\bigr)\leq |s|\,\rho(f)$ for any $v\in \Fq^n$ and,  therefore, we can exchange $S_\lambda ^s$ for $S_f^s$ by creating $|s|\rho(f)$ virtual errors. That is
    $$S_\lambda^s\ket{c+g_1+e_X}=\ket{S_f^s(c+g_1)+e_c+S_\lambda^s e_X},\qquad e_c=L_s(c+g_1),$$ 
    with $\|e_c\|\leq |s|\rho(f)$ and $\|S_\lambda^s e_X\|=\|e_X\|=b$, since $S_\lambda$ is monomial.  Note that the error $e_c$ depends on the ket $\ket{c+g_1}$: it depends only on the coordinates $v_{n-s},\ldots,v_{n-1}$ of $v=c+g_1$ if $s>0$ and on $v_0,\ldots,v_{|s|-1}$ if $s<0$. 
        Now the parity check matrix $H_1$ of the code $C_1$ is applied. Since $S_f^s(c+g_1)\in C_1$,
    $$\sigma_1=H_1\bigl(S_f^s(c+g_1)+e_c+S_\lambda^s e_X\bigr)=H_1(S_\lambda^s e_X)+H_1(e_c).$$
    This process induces the measurement  $H_1(e_c)$, and  the projection operator associated with this measurement is  provided by Lemma~\ref{lem:meas} with $H_J=H_1L_s$, $\sigma=\sigma_1-H_1(S_\lambda^s e_X)$, and $J_s$ being the last or first $|s|$ qudits  
$$Q=\sum_{y: H_J(y_J)=\sigma}\ket{y}\!\bra{y}
=\frac1{q^{m}}\sum_{\beta\in\Fq^{m}}\omega^{-\operatorname{tr}(\beta\cdot\sigma)}\,Z\bigl(\iota_JH_J^{\!\top}\beta\bigr),$$
    this projection consists on $Z$ operators acting on $|s|$ qudits. We interpret these errors as erasures by taking $J_x=\emptyset$ and $J_z=J_s$ in Lemma~\ref{lem:erasures}.  As such, the correction is successful as long as  $2p+|s|\leq d_2-1$. 
    Note that this projected state $e_c$ is constant in the following sense: if $H_1L_s(v)=H_1L_s(v')$, then $L_s(v)-L_s(v')\in C_1$
    has weight at most $2|s|\rho(f)\le d_1-1$ (by the hypothesis $b+|s|\rho(f)\le\floor{(d_1-1)/2}$), so it vanishes. Hence bounded-distance decoding returns the
    fixed vector $S_\lambda^s e_X+e_c$ and subtracting it is a well-defined unitary. By Lemma~\ref{lem:meas}, the
    projection is a linear combination of phase flips supported on $J_s$. 
     
    \item   As $c\in  C_2$ it can be expressed as $\mu  g_2$ for some polynomial $\mu $. After step B1 the central window holds, for every term of the superposition,
    $$S_f^s(c+g_1)=x^s(\mu  g_2+g_1)=x^s\left(\mu  \frac{g_2}{g_1}g_1+g_1\right).$$
    The syndrome of $C_2$ is now measured on the central window. Since $x^s\mu g_2\in C_2$, it equals the syndrome of $x^sg_1$ for every term, so the measurement does not disturb the state, and it determines $x^sg_1\bmod g_2$, that is, $g_1\cdot\bigl(x^s\bmod \frac{g_2}{g_1}\bigr)$. Since $s\in[-a_l,a_r]$ and $a_l+a_r<\text{ord}\left(\frac{g_2}{g_1}\right)$, this determines $s$ uniquely. 
    We then add the virtual error $e_c$ back, so that the central window holds $S_\lambda^s(c+g_1)$, and the alignment is corrected by shifting the frame as in Theorem~\ref{theo:consta1}, which acts as $S_\lambda^{-s}$ on the central window. 
    \item   The final step is the standard one: the padding qudits are deallocated, and the parity check matrix  $H_2$ of $C_2$  and the phase parity check matrix $H_2^\perp$  of the code $C_2^\perp$  are applied on the new central window.  The phase parity check must correct the $p$ phase flips in the central window and the $|s|$ phase-flips generated by measurements. We know which qudits are affected by these phase-flips, so we can interpret them as erasures and apply Lemma~\ref{lem:erasures}.
\end{enumerate}
\end{itemize}
\end{proof}
Let $L_s=S_\lambda^s-S_f^s$ be the map sending $v$ to its virtual error and let
$J_s=\{n-s,\ldots,n-1\}$ if $s>0$ and $J_s=\{0,\ldots,|s|-1\}$ if $s<0$. The proof of
Theorem~\ref{theo:main1} shows that $L_sv$ depends only on $v_{J_s}$, so the syndrome measurement can only
damage the qudits in $J_s$, which are known once $s$ is known. Whether it damages them at all depends on
$$\varepsilon_s=\begin{cases}0,& H_1L_sv=0\ \text{for all } v\in C_2,\\ 1,&\text{otherwise,}\end{cases}$$
and the phase condition of Theorem~\ref{theo:main1} can be sharpened to
$$2p+|s|\,\varepsilon_s\le d_2-1 .$$ We have the following particular cases: 
\begin{enumerate}
\item If $g_1\mid x^n-\lambda$, then the code $C_1$ is also a $\lambda$-constacyclic code, so
$S_\lambda^sC_1=S_f^sC_1=C_1$ and $\varepsilon_s=0$ for every $s$, and the syndrome of the code $C_1$  is not affected by the
virtual error and synchronization costs nothing in phase. This includes the constacyclic case
$\rho(f)=0$.
\item Conversely, for $s=1$ we have $L_1v=v_{n-1}\tilde f_\lambda$, so $\varepsilon_1=0$ if and only if
$\tilde f_\lambda\in C_1$, i.e.\ $g_1\mid\tilde f_\lambda=f-(x^n-\lambda)$, thus \ $g_1\mid x^n-\lambda$.
\item If $f=x^n+ax^i+f_0$ is a  trinomial where $a\neq0$ and $1\le i\le n-1$, then $f-(x^n-\lambda)=ax^i$, so
$\gcd(f,x^n-\lambda)=\gcd(x^n-\lambda,ax^i)=1$, because $x^n-\lambda$ does not vanish at $0$. Hence
$g_1\mid x^n-\lambda$ forces $g_1=1$, and $\varepsilon_1=1$ for every non-trivial marker: the phase cost is
unavoidable for all trinomial codes, in particular for those of Section~\ref{sec:experimental}.
\end{enumerate}
Since the damaged positions are known, they count as erasures and not as errors. For even $d_2$, a
one-position misalignment is therefore free also in phase: $2p+1\le d_2-1$ is equivalent to
$p\le\floor{(d_2-1)/2}$.
Just as in  Theorem~\ref{theo:consta2}, using an asymmetric CSS condition yields a more general result.
\begin{theorem}\label{theo:main2}
    Let $C_1=\langle g_1(x)\rangle$ and $C_3=\langle g_3(x)\rangle$ be two polycyclic codes over the algebra $\Fq[x]/\langle f(x)\rangle$ with $f_0\neq 0$, and $C_2$ a $\mathbb F_q$-linear code. Let  $[n,k_1,d_1]$, $[n ,k_2,d_2]$ and $[n,k_3,d_3]$ be the parameters of $C_1$, $C_2$ and $C_3$ respectively. If $C_2^\perp\subset C_3\subset C_1$ and $a_l+a_r <\text{ord}\left( \frac{g_3}{g_1}\right)$, then there exists an $(a_l,a_r)$-$[[n+a_l+a_r, k_3+k_2-n]]$ quantum synchronizable code that corrects up to $b$ bit-flips, $p$ phase-flips and $s$ misalignments if
    $$
        b+|s|\rho(f)\leq \left\lfloor \frac{d_1-1}{2}\right\rfloor,\quad
       2p+|s|\leq d_2-1,\quad
        s\in [-a_l,a_r].
    $$
\end{theorem}
\begin{proof}
The proof is that of Theorem~\ref{theo:main1}, with step (A1) replaced by step (A1) of Theorem~\ref{theo:consta2}, the syndrome of $C_3$ in place of that of $C_2$ in step (B2), and Lemma~\ref{lem:erasures} applied to the pair $C_2^\perp\subseteq C_3$ in step (B3).
\end{proof}

\section{Quasi-polycyclic QSCs}\label{sec:general}
We can assign a family of projection operators $\pi_i$ for $i=1,\ldots,\ell$ to each $\Fq[x]$ module $(\Fp)^\ell$ that project the polynomial vector $c$ into each one of its polycyclic blocks:
\begin{align*}
    \pi_i:(\Fp)^\ell&\longrightarrow \Fp,\\
    (c_1(x),\ldots,c_\ell(x))&\longmapsto c_i(x).
\end{align*}
It is clear that, for
     $C$  a quasi-polycyclic code over the module $\mathcal A^\ell$, $\pi_i(C)$ is a polycyclic code over the algebra $\mathcal A$.
As mentioned before,  
misalignment errors differ from Pauli errors: the former are global while the latter are local. In the quasi-polycyclic case we will  use this property to our advantage. If we were to detect a misalignment $S^s$ in the polycyclic block $\pi_i(C)$ through Theorem~\ref{theo:main1}, and correct it by applying $S^{-s}$, the misalignment would be corrected in the whole word, not just in the subword.
Note that
it is possible to apply the CSS construction using annihilator duals \cite{bag2026annihilatortwistedeuclideanduality,polydual}, thus, this construction will help search for quantum quasi-polycyclic codes in later sections.
\begin{theorem}\label{theo:cssanh}
    Let $C_1$ and $C_2$ be two quasi-polycyclic codes over $\mathcal A^\ell$ with $f_0\neq 0$ and parameters $[N,k_1]$ and $[N,k_2]$, respectively. If $C_2^\circ\subset C_1$, there exists an $[[N,k_1+k_2-N]]$ quantum code that corrects up to  $\left\lfloor \frac{d_X-1}{2}\right\rfloor$ bit-flip errors and $\left\lfloor \frac{d_Z-1}{2}\right\rfloor$ phase-flip errors, where $d_X$ and $d_Z$ are the minimal distances of $\Psi(C_1)$ and $\Psi(C_2)\cdot \mathcal G$, respectively.
\end{theorem}
\begin{proof}
     Theorem \ref{teo:twist}  states  $(\Psi(C_2)\cdot \mathcal G)^\perp=\Psi(C_2^\circ)$. So $C_2^\circ\subset C_1$ becomes $(\Psi(C_2)\cdot \mathcal G)^\perp\subset \Psi(C_1)$.  Now we can apply Theorem \ref{theo:css} to this new chain of codes to obtain a $[[N,k_1+k_2-N]]$ quantum code. Note that $\dim (\Psi(C_2)\cdot \mathcal G)=\dim(\Psi(C_2))$, indeed $f_0\neq 0$ and therefore $\mathcal G$ is invertible. 
\end{proof}

  \begin{theorem}\label{theo:main3}
    Let $C_2$ be a quasi-polycyclic code over the module $\mathcal A^\ell$ with $f_0\neq 0$ and parameters $[N,k_2,d_2]$, $N=\ell n$, such that $C_2^\circ\subset C_2$. Let $C_1=\langle g_1(x)\rangle$ be a polycyclic code over $\mathcal A$ with parameters $[n,k_1,d_1]$. Let $i\in[1,\ell]$ be an integer so that $\pi_i(C_2)=\langle g_2(x)\rangle $ and  $\pi_i(C_2)\subset C_1$.  If $a_l+a_r <\text{ord}\left( \frac{g_2}{g_1}\right)$, then there exists an $(a_l,a_r)$-$[[N^*, 2k_2-N]]$ quantum synchronizable code where 
$$N^*=N+a_l+a_r,$$
that corrects up to $b$ bit-flips, $b_i$ bit flips on $\pi_i(C_2)$, $p$ phase-flips and $s$ misalignments if
  
    $$
        b_i+|s|\rho(f)\leq \left\lfloor \frac{d_1-1}{2}\right\rfloor,\quad 
        b\leq \floor{\frac{d_2-1}{2}},\quad
        2p+|s|\leq d_Z-1,\quad
        s\in [-a_l,a_r],
    $$
 where $d_2=d(\Psi(C_2))$ is the bit-flip distance and $d_Z$ is the minimal distance of $\Psi(C_2)\cdot \mathcal G$.
 \end{theorem}
    \begin{proof}
As usual, we follow the schema as in Theorem~\ref{theo:main1}.
\begin{itemize}
    \item[A]  Encoding.
    \begin{enumerate}
        \item    We apply the annihilator CSS construction to get a sum of kets  $\sum_i \alpha_i\ket {v_i+C_2^\circ}$. As in Theorem~\ref{theo:main1}, we keep track of the whole superposition, and for each of its terms $\ket c$, $c\in C_2$, we focus on the coordinates $\pi_i(c)=c_i\in \pi_i(C_2)$.
    \item To encode the alignment information, the generator $g_1$ is added to the state  $\ket{ c_i+g_1}$ with $ c_i+g_1\in C_1.$ 
    \item We add twisted padding to the $i$-th component only. In total, $N^*=N+a_l+a_r$ qudits are sent:
    $$\ket {c_1}\ldots \ket{c_{i-1}}\ket{c_i+g_1}_{(a_l,a_r)}^\lambda\ket{c_{i+1}}\ldots \ket{c_\ell}. $$ 
    If $s\in[-a_l,a_r]$, Lemma~\ref{lem:twisted} shows that the window read in the $i$-th component is
    $S_\lambda^{s}(c_i+g_1)$. The padding of the $i$-th component guarantees that this window lies inside the
    transmitted block for every $s\in[-a_l,a_r]$. Once $s$ is known, the receiver shifts its frame as in
    Theorem~\ref{theo:consta1}, so every component is re-addressed without loss.
    \end{enumerate}
     \item[B]  Decoding.
    \begin{enumerate}
    \item The message is received along with $p$ phase-flips and $b$ bit flips, of which $b_i$ occur on the $i$-th component. We exchange $S_\lambda$ for $S_f$ and correct up to $\left\lfloor \frac{d_1-1}{2}\right\rfloor$  bit-flips on  the $i$-component $(\ket{c_i+g_1}_{(a_l,a_r)}^\lambda)$ like in Theorem~\ref{theo:main1}.
    \item   We detect and correct the misalignment on the $i$-component as in Theorem~\ref{theo:main1}, thus the misalignment is corrected on the whole word. 
    \item  Bit-flip errors and phase-flip errors are corrected on the whole word using the phase-flip $H_2^\circ$ and bit-flip $H_2$ parity check matrices. The phase-flips generated by measurement are taken as erasures, as their position is known.
    \end{enumerate}
\end{itemize}
\end{proof}

Quasi-polycyclic synchronizable codes carry two bit-flip conditions:
$b_i+|s|\rho(f)\le\floor{\frac{d_1-1}{2}}$ on the synchronization component and
$b\le\floor{\frac{d_2-1}{2}}$ on the whole word, and both are needed for the guarantee. In particular,
bursts of bit flips concentrated on the $i$-th component are governed by $d_1$, which is usually much
smaller than $d_2$. For $\rho(f)=0$, as in the quasi-constacyclic case below, the first condition becomes
$b_i\le\floor{\frac{d_1-1}{2}}$.

In the case in which the codes are quasi-constacyclic, we get the following result. 

    \begin{theorem}\label{theo:main3consta}
    Let $\mathcal A=\Fq[x]/\langle x^n-\lambda\rangle$, let $C_2$ be a quasi-constacyclic code over the module $\mathcal A^\ell$ with parameters $[N,k_2,d_2]$, $N=\ell n$, such that $C_2^\circ\subset C_2$. Let $C_1=\langle g_1(x)\rangle$ be a $\lambda$-constacyclic code over $\mathcal A$ with parameters $[n,k_1,d_1]$. Let $i\in[1,\ell]$ be an integer so that $\pi_i(C_2)=\langle g_2(x)\rangle $ and  $\pi_i(C_2)\subset C_1$.  If $a_l+a_r <\text{ord}\left( \frac{g_2}{g_1}\right)$, then there exists an $(a_l,a_r)$-$[[N^*, 2k_2-N]]$ quantum synchronizable code where 
$$N^*=N+a_l+a_r,$$
that corrects up to $b$ bit-flips, $b_i$ bit flips on $\pi_i(C_2)$, $p$ phase-flips and $s$ misalignments if
  
    $$
        b_i\leq \left\lfloor \frac{d_1-1}{2}\right\rfloor,\quad 
        b\leq \floor{\frac{d_2-1}{2}},\quad
        p\leq \left\lfloor \frac{d_Z-1}{2}\right\rfloor,\quad
        s\in [-a_l,a_r],
    $$
 where $d_2=d(\Psi(C_2))$ is the bit-flip distance and $d_Z$ is the minimal distance of $\Psi(C_2)\cdot \mathcal G$.
 \end{theorem}

\section{Experimental data}\label{sec:experimental}

This section collects the codes found by computer search using
Magma~\cite{Magma}, version V2.29-7, on  a Linux
machine with an AMD EPYC 7F52 processor (16 cores, 32 threads) and 512~GB of memory; each search
ran on a single core. The searches took from a few seconds to several hours each, depending on the
length and the number of pairs $(g,h)$ explored. A script that rebuilds every code from the stored
data and checks its parameters runs in about half a minute for the tables of index~$1$ and for
Tables~\ref{tab:exp:binomcap} and~\ref{tab:exp:q5}; for Table~\ref{tab:exp:q7} it takes a few
seconds for each binomial code and between about $6$ and $14$ hours for each trinomial code, almost
all of it spent computing the exact distances $d_X=d_Z=7$. The data and the scripts are
available at \url{https://github.com/edgarmartinezmoro-web/qsc-magma}.  Throughout, $d_X$ and $d_Z$ denote the classical
bit-flip and phase-flip distances that appear in the correction conditions of our theorems; the
true quantum distance may be larger if the codes are degenerate.
We will compare four families of codes in this section.
\begin{enumerate}
\item cyclic and constacyclic codes ($\rho=0$).
\item polycyclic codes of index~$1$ associated with trinomials ($\rho=1$), from the codes
in~\cite{trinomials};
\item quasi-cyclic and quasi-constacyclic codes of index~$2$ ($\rho=0$), including the family
in~\cite{DuMaLiu2023};
\item quasi-polycyclic codes of index~$2$ over repeated-root trinomials ($\rho=1$).
\end{enumerate}

To read  the tables we must remark that, by Theorems~\ref{theo:main1}-\ref{theo:main3}, a misalignment
of $s$ positions costs $|s|\rho(f)$ bit flips in the synchronization code, of distance $d_1$, and,
when the syndrome of that code depends on the transmitted state, $|s|$ phase flips at known
positions, which are corrected as erasures; see Lemma~\ref{lem:erasures}. With $b_1$ bit flips in
the synchronization block, $b$ bit flips in total and $p$ phase flips, the conditions are
\begin{equation}\label{eq:exp:conditions}
b_1+|s|\rho(f)\le\floor{\tfrac{d_1-1}{2}},\qquad b\le\floor{\tfrac{d_X-1}{2}},\qquad
2p+|s|\varepsilon_s\le d_Z-1,
\end{equation}
where $\varepsilon_s\in\{0,1\}$ is $0$ when the syndrome does not detect the virtual error, this is always
the case for $\rho=0$. For all the trinomial codes below $\gcd(f,x^{\deg f}+f_0)=1$, so $\varepsilon_1=1$, and we take $\varepsilon_s=1$ for every $s\neq0$, which is a safe bound.
Two consequences are used repeatedly in the following sections: the nominal window $a_l+a_r<\operatorname{ord}(g/g_1)$ is only
useful up to  $|s|\rho(f)\le\floor{(d_1-1)/2}$ (even with $b_1=0$), so $d_1$ matters as much as the window; and for even $d_Z$
a one-position misalignment costs nothing in phase.

 \subsection{Index one: QSCs from trinomial codes}\label{sec:exp:index1}

In~\cite{trinomials}, quantum codes are built from a polycyclic code $C_1=\langle g_1\rangle$ associated
with a trinomial $f$ and a divisor $h_2\mid f$, under the annihilator condition
$\langle f/h_2\rangle\subseteq C_1$. By Theorem~\ref{teo:twist} the Euclidean pair behind it is
$C_2^{\perp}=\langle f/h_2\rangle\subseteq C_1$ with $C_2=\langle f/h_2\rangle^{\perp}$, a linear code which is
not polycyclic, so these codes fit the conditions of the asymmetric case stated in Theorem~\ref{theo:main2}. For every monic polynomial
$t\mid f/(h_2g_1)$, $t\neq1$, take the code $C_3=\langle g_1t\rangle$; then $C_2^{\perp}\subseteq C_3\subseteq C_1$ and we
obtain an $(a_l,a_r)$-$[[n+a_l+a_r,K']]_q$ QSC with $K'=k_1+k_2-n-\deg t$ and window
$a_l+a_r<\operatorname{ord}(t)$, where $d_1=d(C_1)$ and $d_2=d(C_2)$ do not depend on the chosen polynomial $t$.

As competitors, we took every QSC obtained in the same way from constacyclic codes, by an
exhaustive search over all $\lambda\in\Fq^{*}$ and all chains $g_1'\mid g_3'\mid u'\mid x^n-\lambda$,
with $C_2'=\langle u'\rangle^{\perp}$ and $\operatorname{ord}(g_3'/g_1')\ge3$. In all four cases in Table~\ref{tab:exp:aly}, no cyclic or constacyclic QSC of the same length and dimension reaches the pair
$(d_1,d_2)$ of the trinomial QSC. For the codes $[[60,32,5]]_3$ ($f=x^{60}-2x^6-2$, $(d_1,d_2)=(5,6)$)
and $[[72,48,5]]_3$ ($f=x^{72}-x^2-1$, $(d_1,d_2)=(5,5)$) in~\cite{trinomials} we obtain the families of
Table~\ref{tab:exp:aly5}; an exhaustive search of constacyclic competitors is not feasible at these
lengths.

\begin{table}[h]
\centering
\small
\resizebox{\textwidth}{!}{%
\begin{tabular}{cccccc}
\toprule
code of~\cite{trinomials} & $f$ & $t$ & polycyclic QSC & $(d_1,d_2)$ & best (consta)cyclic\\
\midrule
$[[10,2,4]]_{13}$ & $x^{10}-2x-1$ & $[21]$ & $[[10+a_l+a_r,1]]$, $a_l+a_r<12$ & $(4,4)$ & $(2,2)$\\
$[[16,10,3]]_{13}$ & $x^{16}-2x-4$ & $[81]$ & $[[16+a_l+a_r,9]]$, $a_l+a_r<4$ & $(3,3)$ & $(3,2)$ or $(2,3)$\\
$[[27,17,4]]_{7}$ & $x^{27}-2x^3-3$ & $[2001]$ & $[[27+a_l+a_r,14]]$, $a_l+a_r<18$ & $(4,4)$ & $(2,2)$\\
$[[27,17,4]]_{7}$ & $x^{27}-2x^3-3$ & $[243251]$ & $[[27+a_l+a_r,12]]$, $a_l+a_r<16806$ & $(4,4)$ & $(3,2)$ or $(2,3)$\\
\bottomrule
\end{tabular}}
\caption{QSCs from trinomial codes of~\cite{trinomials} and the Pareto-optimal pairs $(d_1,d_2)$
among all constacyclic QSCs of the same length and dimension. A string $[c_0c_1\cdots c_r]$ denotes
$c_0+c_1x+\cdots+c_rx^r$.}
\label{tab:exp:aly}
\end{table}

\begin{table}[h]
\centering
\small
\begin{tabular}{ccccc}
\toprule
$n$ & $t$ & $K'$ & window $a_l+a_r<\operatorname{ord}(t)$ & $(d_1,d_2)$\\
\midrule
$60$ & $[11]$ & $31$ & $2$ & $(5,6)$\\
$60$ & $[121]$ & $30$ & $6$ & $(5,6)$\\
$60$ & $[1021]$ & $29$ & $26$ & $(5,6)$\\
$60$ & $[120211]$ & $27$ & $78$ & $(5,6)$\\
$60$ & $[1211021]$ & $26$ & $182$ & $(5,6)$\\
$60$ & $[101220001]$ & $24$ & $546$ & $(5,6)$\\
\midrule
$72$ & $[10101020101]$ & $38$ & $484$ & $(5,5)$\\
$72$ & (degree $19$) & $29$ & $\approx 1.16\cdot 10^{9}$ & $(5,5)$\\
\bottomrule
\end{tabular}
\caption{QSC families $(a_l,a_r)$-$[[n+a_l+a_r,K']]_3$ from the codes $[[60,32,5]]_3$ and
$[[72,48,5]]_3$ of~\cite{trinomials}. The polynomials $t$ are written as in Table~\ref{tab:exp:aly}.}
\label{tab:exp:aly5}
\end{table}

By Equation~\eqref{eq:exp:conditions}, the codes of Table~\ref{tab:exp:aly}   tolerate $|s|\le1$ since $d_1\in\{3,4\}$
and those of Table~\ref{tab:exp:aly5} tolerate $|s|\le2$ since $d_1=5$; therefore larger windows add no
synchronization capability. For $n=60$ the smallest useful member is the $(2,2)$-$[[64,30]]_3$ code
where $t=[121]=(x+1)^2$, of order $6$; with $|s|=2$ it still corrects one phase flip ($2p+2\le5$), and with
$|s|\le1$ two. Constacyclic competitors, with smaller distances, synchronize over their whole
window at no cost. The trinomial codes are therefore preferable when misalignments of more
than one or two positions are rare, as in the channel model of Section~\ref{sec:channel} with
small $p_s$.

These trinomial codes have higher distances $d_1, d_2$ than their constacyclic counterparts $\tilde d_1,\tilde d_2$, but they also suffer virtual errors. When does the distance gained pay off for the virtual errors generated? If $n,k,a_l,a_r$ are fixed the answer will depend on the parameters of the channel $p_x,p_z,p_l,p_r$. For simplicity, we will study the symmetric case in which $p_s:=p_l=p_r$. We will denote $N=n+a_l+a_r$. We recall the probability distributions of the different errors. 
\begin{align*}
   P(b=j)=&\binom{ N}{j}p_x^j(1-p_x)^{N-j},\\
   P(p=j)=&\binom{ N}{j}p_z^j(1-p_z)^{N-j},\\
   P(s=j)=&\frac{1-p_s}{1+p_s}\,p_s^{|j|}.
\end{align*}
We now define constacyclic and polycyclic thresholds:
\begin{align*}
    \tilde T_b&= \left\lfloor \frac{\tilde d_1-1}{2} \right\rfloor, \qquad \tilde T_z = \left\lfloor \frac{\tilde d_2-1}{2} \right\rfloor,\\
      T_b&= \left\lfloor \frac{d_1-1}{2} \right\rfloor, \qquad T_z = d_2-1.\\
\end{align*}

Then the probability of success of the constacyclic code is:
$$ \mathcal{P}_{\mathrm{succ}}^{(c)} = \mathrm{P}(b \leq \tilde T_b) \cdot \mathrm{P}(p \leq \tilde T_z) \cdot \mathrm{P}(-a_l \leq s \leq a_r),$$
and the probability of success of the polycyclic code is:
$$\mathcal{P}_{\mathrm{succ}}^{(p)} = \sum_{j=-a_l}^{a_r} \mathrm{P}(s = j) \cdot \mathrm{P}\big(b \leq T_b - |j|\rho(f)\big) \cdot \mathrm{P}\left(p \leq \left\lfloor \frac{T_z - |j|}{2} \right\rfloor\right).$$

\begin{figure}
    \centering
    \begin{subfigure}{0.49\linewidth}
        \centering
        \includegraphics[width=\linewidth]{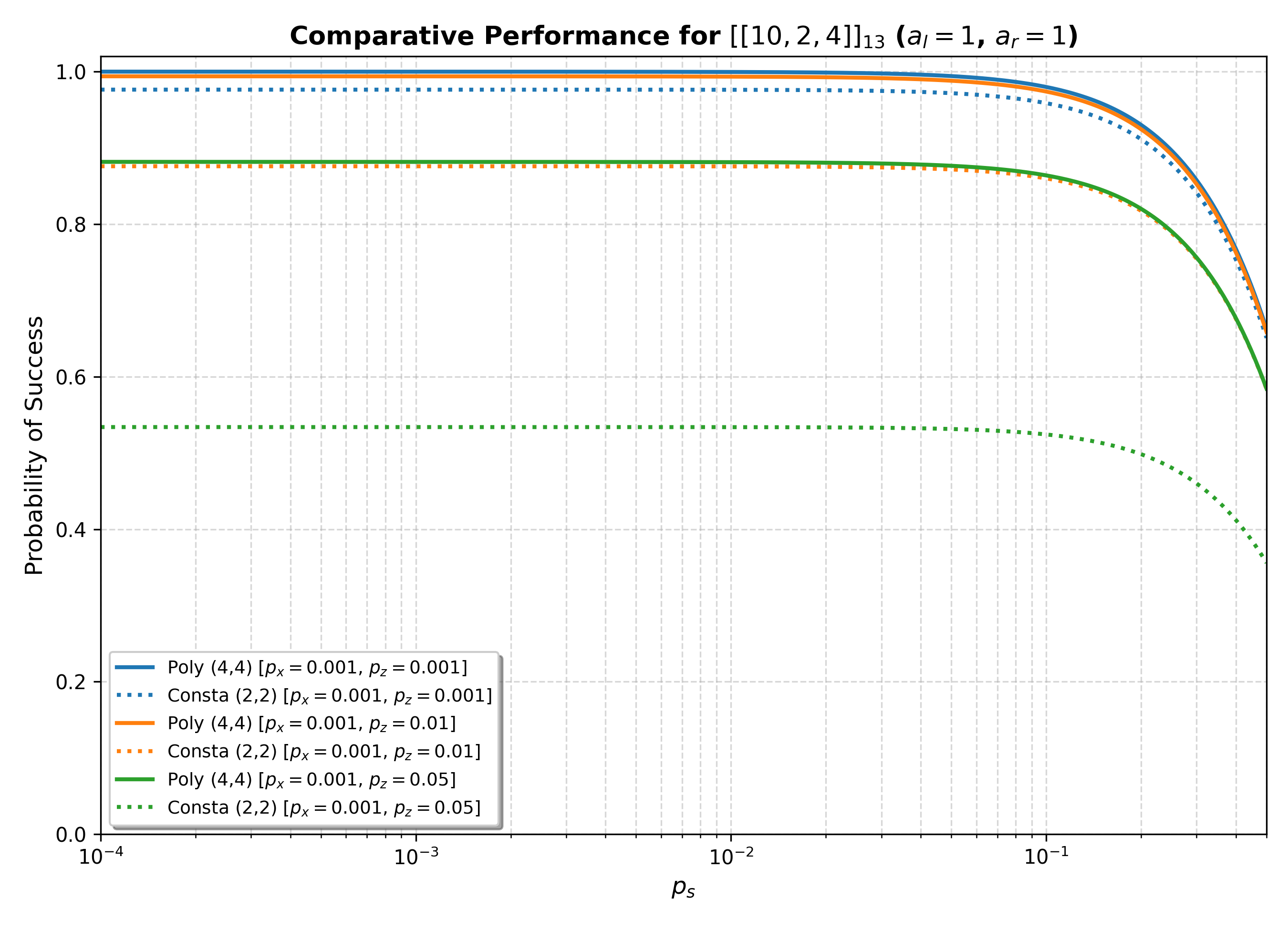}
        \caption{$[[10+a_l+a_r,1]]_{13}$, $(4,4)$ vs.\ $(2,2)$}
        \label{fig:codes1a}
    \end{subfigure}\hfill
    \begin{subfigure}{0.49\linewidth}
        \centering
        \includegraphics[width=\linewidth]{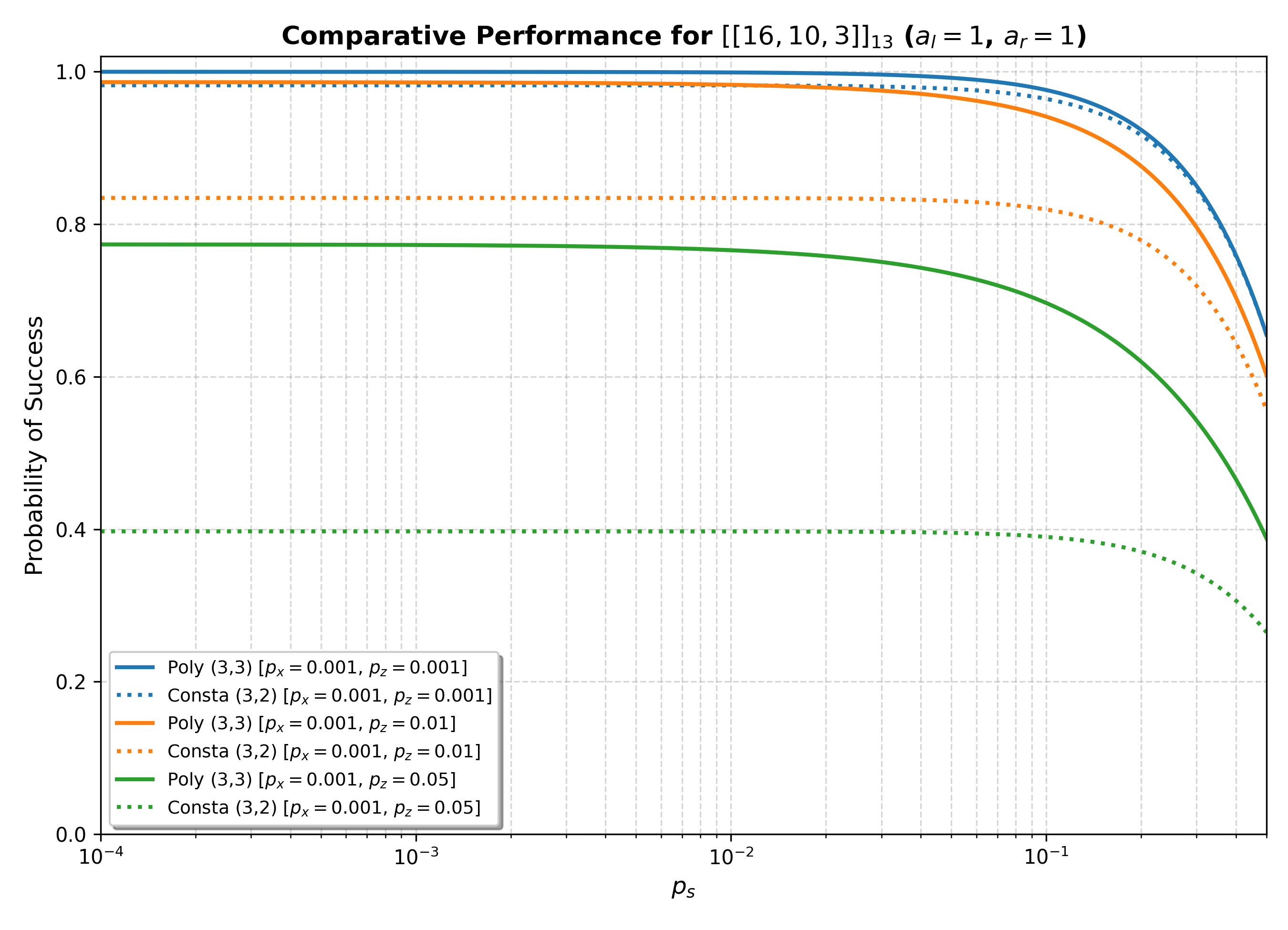}
        \caption{$[[16+a_l+a_r,9]]_{13}$, $(3,3)$ vs.\ $(3,2)$}
        \label{fig:codes1b}
    \end{subfigure}

    \medskip
    \begin{subfigure}{0.49\linewidth}
        \centering
        \includegraphics[width=\linewidth]{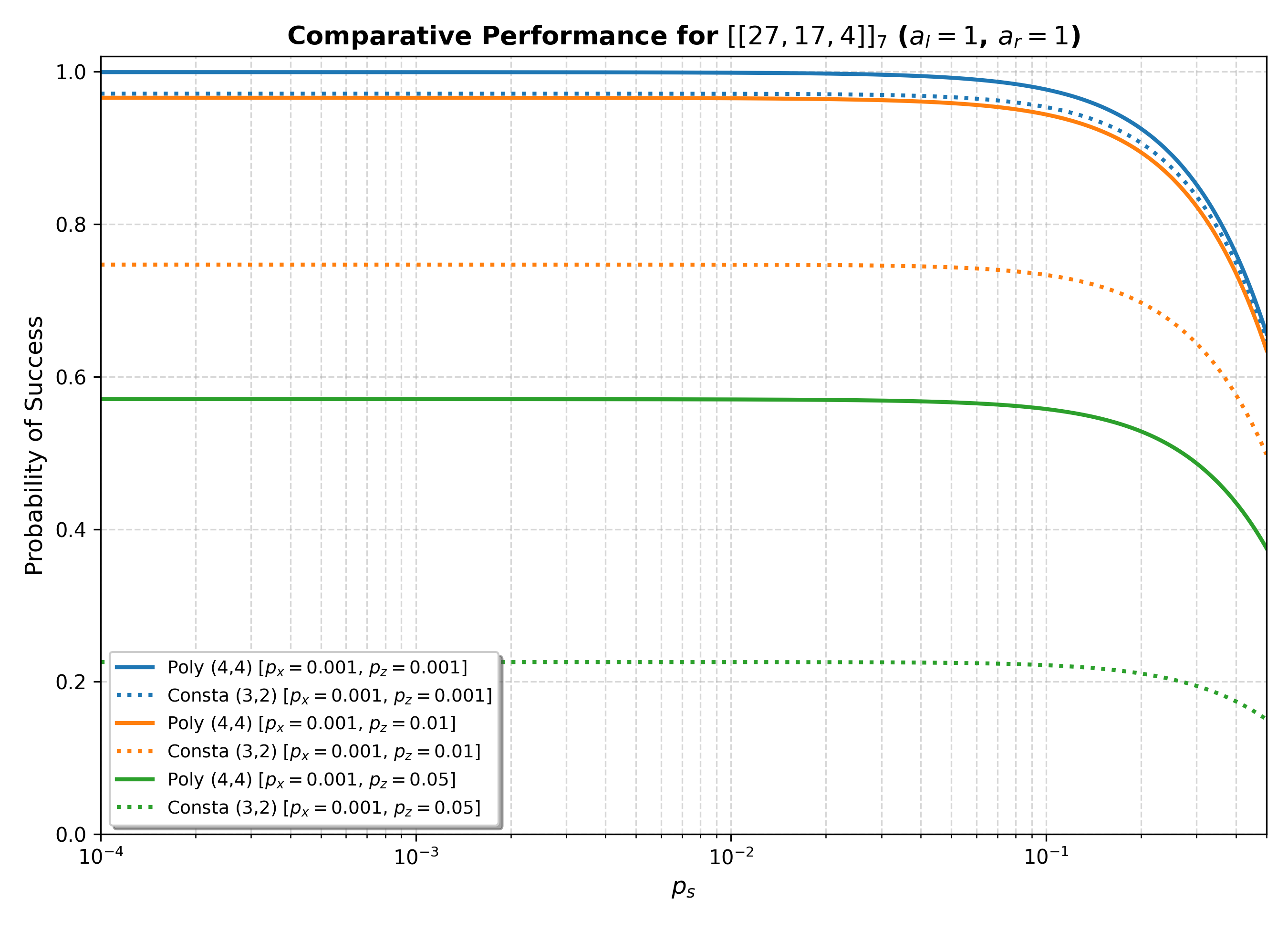}
        \caption{$[[27+a_l+a_r,12]]_{7}$, $(4,4)$ vs.\ $(3,2)$}
        \label{fig:codes1c}
    \end{subfigure}
    \caption{Probability of success of three of the QSCs of Table~\ref{tab:exp:aly} ($a_l=a_r=1$) in several error regimes, compared with the best constacyclic competitor of the same length and dimension.}
    \label{fig:codes1}
\end{figure}

Figure~\ref{fig:codes1} compares both probabilities for three of the QSCs of Table~\ref{tab:exp:aly}, with $a_l=a_r=1$, against the best constacyclic competitor found by the exhaustive search: $[[10+a_l+a_r,1]]_{13}$ with $(d_1,d_2)=(4,4)$ against $(2,2)$, $[[16+a_l+a_r,9]]_{13}$ with $(3,3)$ against $(3,2)$, and $[[27+a_l+a_r,12]]_7$ ($t=[243251]$ in the notation of Table~\ref{tab:exp:aly}) with $(4,4)$ against $(3,2)$. In all three cases the trinomial QSC has a probability of success at least as high as its competitor in every error regime considered, strictly higher except at the largest values of $p_s$, and its gains are larger the more asymmetric the noise is. Since these competitors are actual codes, these are the comparisons on which our conclusions rest.

\begin{figure}
    \centering
    \begin{subfigure}{0.49\linewidth}
        \centering
        \includegraphics[width=\linewidth]{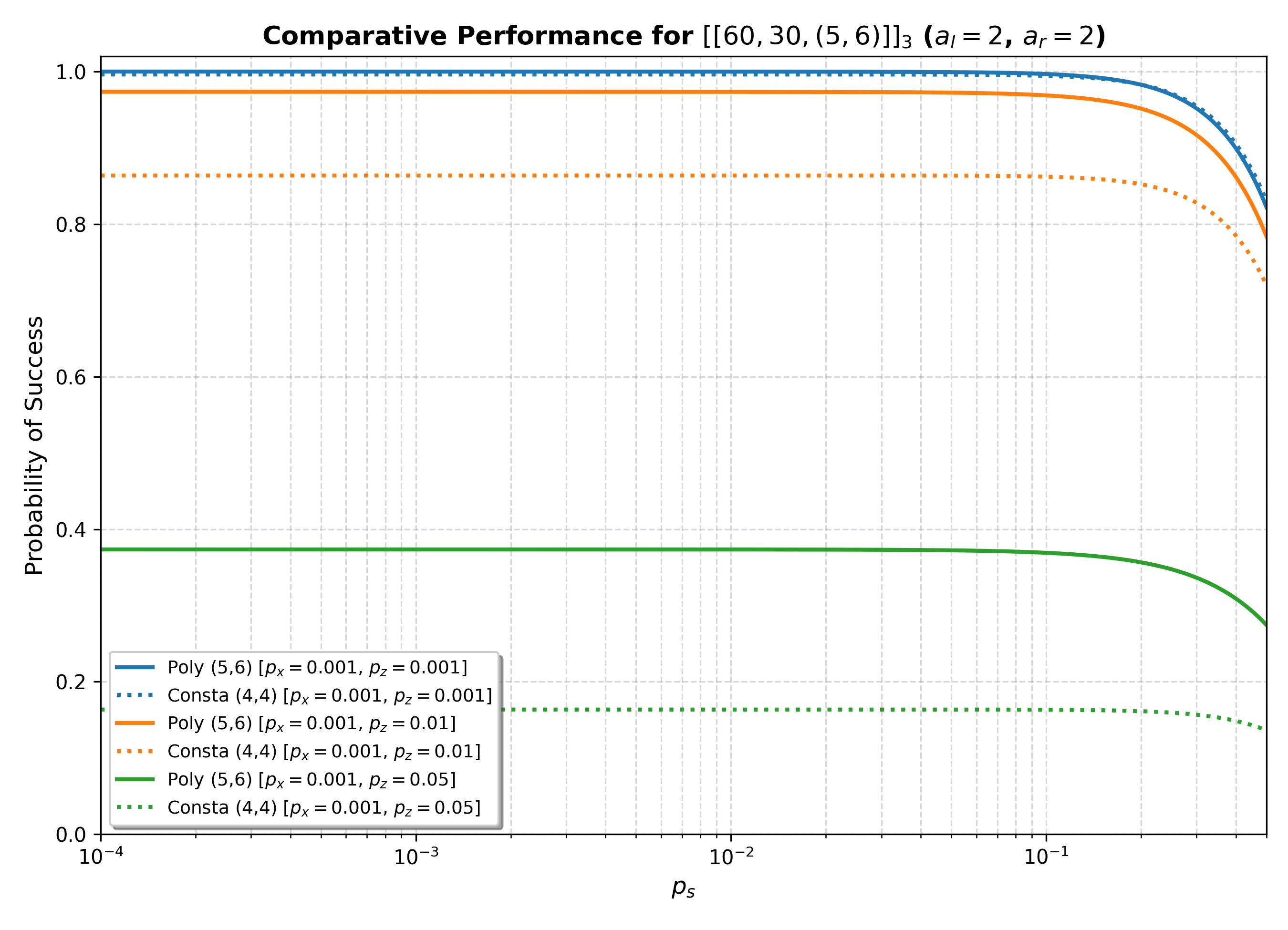}
        \caption{vs.\ hypothetical $(3,3)$ or $(4,4)$}
        \label{fig:codes2a}
    \end{subfigure}\hfill
    \begin{subfigure}{0.49\linewidth}
        \centering
        \includegraphics[width=\linewidth]{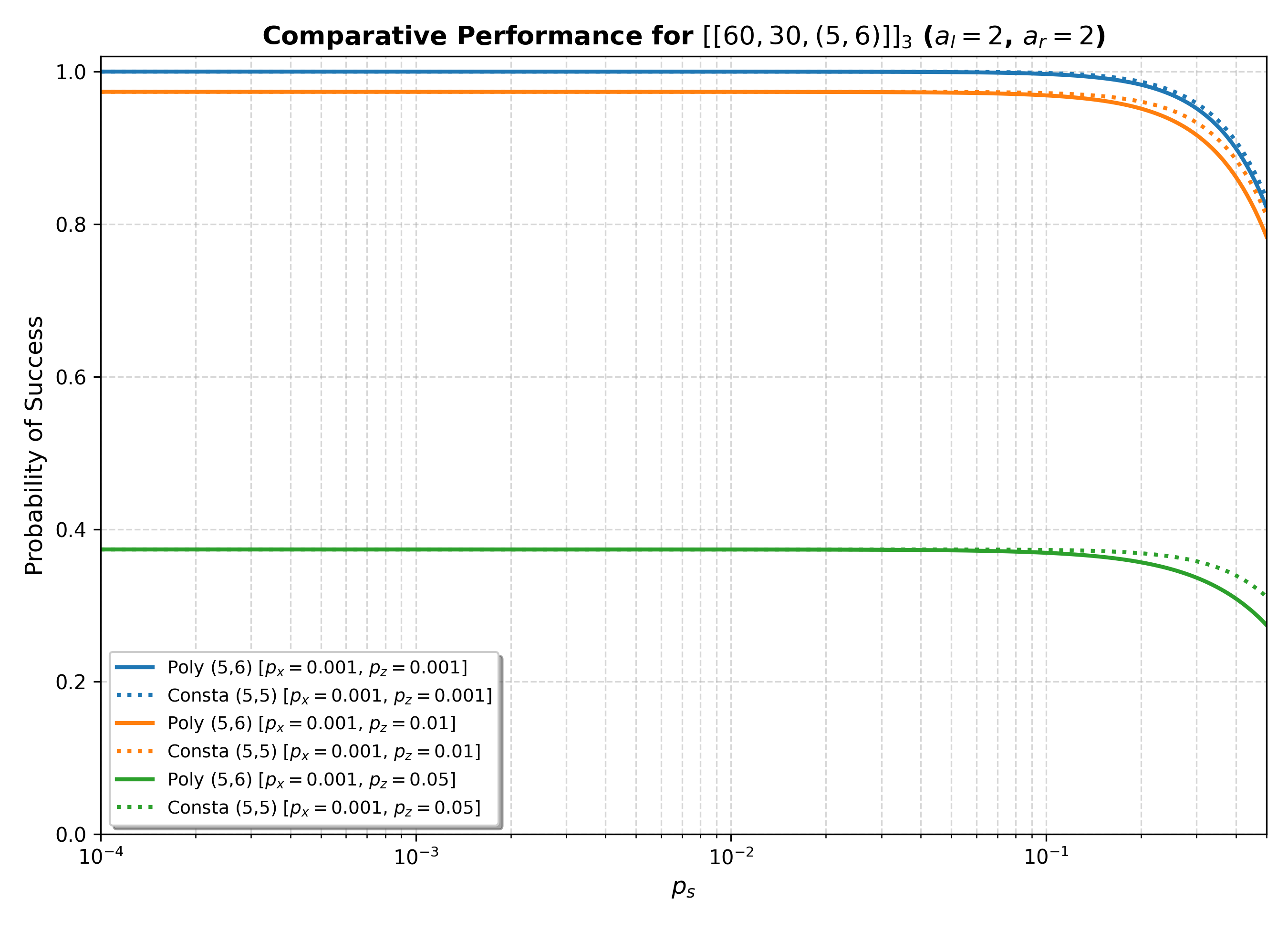}
        \caption{vs.\ hypothetical $(5,5)$}
        \label{fig:codes2b}
    \end{subfigure}
    \caption{Probability of success of the QSC $[[60+a_l+a_r,30]]_3$, $(d_1,d_2)=(5,6)$, with $a_l=a_r=2$, compared with hypothetical constacyclic codes. A $(3,3)$ and a $(4,4)$ competitor give the same curves.}
    \label{fig:codes2}
\end{figure}

\begin{figure}
    \centering
    \includegraphics[width=0.5\linewidth]{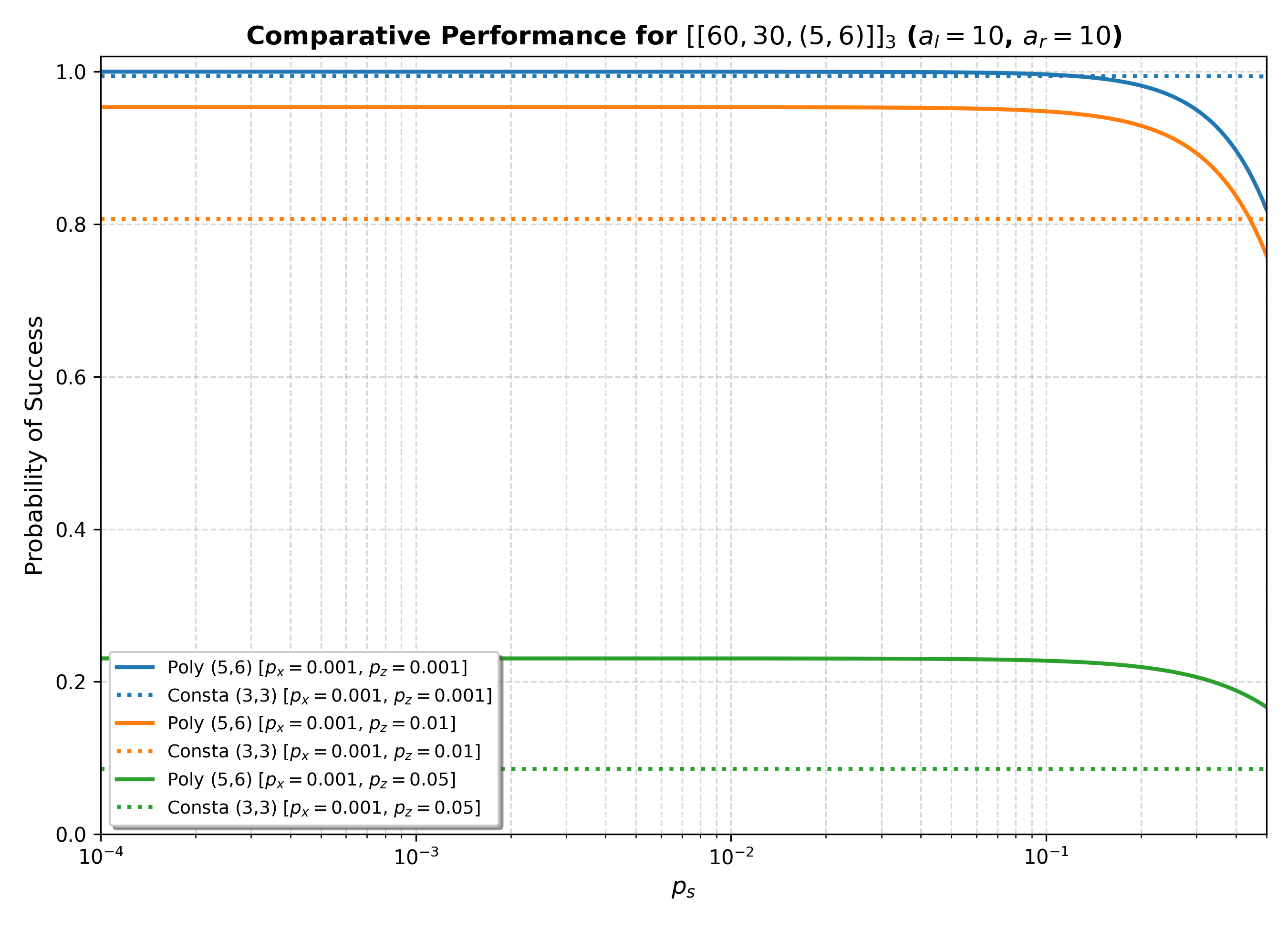}
    \caption{Probability of success of the QSC $[[60+a_l+a_r,30]]_3$, $(d_1,d_2)=(5,6)$, in several error regimes, compared with a hypothetical $(3,3)$ constacyclic code, for a large synchronization window ($a_l=a_r=10$).}
    \label{fig:codes3}
\end{figure}

For the QSC $[[60+a_l+a_r,30]]_3$ of Table~\ref{tab:exp:aly5}, with $(d_1,d_2)=(5,6)$, an exhaustive search of constacyclic competitors is not feasible, and in Figures~\ref{fig:codes2}--\ref{fig:codes3} the competitor is a \emph{hypothetical} constacyclic code with the stated pair $(\tilde d_1,\tilde d_2)$; these pairs are assumptions, not computed codes. The outcome depends on that assumption. Against $(3,3)$ or $(4,4)$, which give identical curves because $\floor{(3-1)/2}=\floor{(4-1)/2}$ (Figure~\ref{fig:codes2a}), the trinomial QSC is better except for very large $p_s$ combined with low phase noise. Against a $(5,5)$ constacyclic code (Figure~\ref{fig:codes2b}) it is never better: both coincide for small $p_s$, since they correct the same number of errors when $s=0$, and the constacyclic code wins for large $p_s$, since it synchronizes at no cost.
Figure~\ref{fig:codes3} shows what happens with large synchronization windows. These windows exceed the misalignment correction capability of the polycyclic code, which cannot correct high enough $|s|$. Even so, for most values of $p_s$ the trinomial codes still surpass their constacyclic analogues. Only for high $p_s$ ($p_s>0.1$) is the constacyclic code better. A possible explanation is that given the error model in Section~\ref{sec:channel}  the probability of $s$ misalignments decreases exponentially in $|s|$.

\subsection{Index two: quasi-cyclic versus quasi-polycyclic codes}\label{sec:exp:qp}

 Let $\mathcal A=\Fq[x]/\langle f\rangle$ with $\deg f=m$, and for monic $g,h\mid f$ and
$k\in\mathcal A$ let $C=\langle(g,kg),(0,h)\rangle\subseteq\mathcal A^{2}$. This is the family in~\cite{Galindo2018} used in~\cite{DuMaLiu2023} with $f=x^m-1$ and Euclidean
duality. We use instead the annihilator duality and the conversion by $\mathcal G$, which works for any $f$.

\begin{lemma}\label{lem:exp:annGalindo} $\quad$
\begin{enumerate}
\item $C^{\circ}$ is generated by $\bigl(-kf/h,\ f/h\bigr)$ and $\bigl(f/g,\ 0\bigr)$, and
$\dim_{\Fq}C=2m-\deg g-\deg h$.
\item If $g\mid h$ and $h^{2}\mid f$, then $C^{\circ}\subseteq C$.
\end{enumerate}
\end{lemma}
\begin{proof}
\textit{(1)} The four products of these vectors with $(g,kg)$ and $(0,h)$ are multiples of $f$, so both
vectors lie in $C^{\circ}$. Their generator matrix is triangular with diagonal $f/h$, $f/g$, so they
span a module of dimension $\deg g+\deg h=2m-\dim C=\dim C^{\circ}$.
\textit{(2)} Since $gh\mid h^2\mid f$,
$$\Bigl(-k\tfrac fh,\tfrac fh\Bigr)=-\tfrac{f}{gh}\,k\,(g,kg)+\tfrac{f(1+k^{2})}{h^{2}}\,(0,h),\qquad
\Bigl(\tfrac fg,0\Bigr)=\tfrac{f}{g^{2}}\,(g,kg)-\tfrac{kf}{gh}\,(0,h),$$
with polynomial coefficients.
\end{proof}

By Theorem~\ref{teo:twist}, $C^{\circ}\subseteq C$ gives $\bigl(\Psi(C)\mathcal G\bigr)^{\perp}=\Psi(C^{\circ})\subseteq\Psi(C)$,
so $\bigl(\Psi(C),\Psi(C)\mathcal G\bigr)$ is a Euclidean CSS pair in physical coordinates, with
$$N=2m,\qquad K=2(m-\deg g-\deg h),\qquad d_X=d(\Psi(C)),\qquad d_Z=d(\Psi(C)\mathcal G).$$
We synchronize on the first block, $\pi_1(C)=\langle g\rangle$, with a marker code $\langle g_1\rangle$, $g_1\mid g$,
$g_1\neq g$, of distance $d_1$ and window $a_l+a_r<\operatorname{ord}(g/g_1)$, as in Theorem~\ref{theo:main3}.
The transmitted length is $N^{*}=2m+a_l+a_r$.

 The family contains the one-block subcode $\{0\}\times\langle h\rangle$,
which caps its distance independently of $k$ and of the length, thus providing a structural bound.

\begin{proposition}\label{prop:exp:cap}
If $(0,h)\in C$, then $d_X\le d(\langle h\rangle)$. Moreover, if $f=x^m-\lambda$, then $d_Z\le d(\langle h\rangle)$.
For $f=x^{rp^{s}}-\lambda$ with $p$ odd, $\gcd(r,p)=1$, $s\ge1$, and $h^2\mid f$, this gives
$$\min(d_X,d_Z)\le\begin{cases}\frac{(p+1)}{2}& s=1,\\ \frac{(p+3)}{2}& s\ge2,\end{cases}$$
for every $g$, $k$, $r$ and every length.
\end{proposition}
\begin{proof}
$\mathcal A\cdot(0,h)=\{0\}\times\langle h\rangle\subseteq C$; for $f=x^m-\lambda$ the Gram matrix $G_{\mathcal A}$ is monomial,
which gives the claim for $d_Z$. For the binomial, write $\lambda=\mu^{p^{s}}$, so $f=(x^r-\mu)^{p^s}$ and
$h\mid(x^r-\mu)^{j}$ with $j=\floor{p^s/2}$. The substitution $y=x^r$ embeds the repeated-root
constacyclic code $\langle(y-\mu)^j\rangle$ of length $p^s$ into $\langle h\rangle$ preserving weights, and that code
is monomially equivalent to $\langle(y-1)^j\rangle$, whose distance is $\beta+1$ for
$(\beta-1)p^{s-1}<j\le\beta p^{s-1}$ by~\cite{Dinh2008}. For $j=(p^s-1)/2$ this is $(p+1)/2$ if $s=1$ and
$(p+3)/2$ if $s\ge2$.
\end{proof}

Proposition~\ref{prop:exp:cap} explains the distances $4,5,7$ in~\cite{DuMaLiu2023} for $p=5,7,11$,
which do not grow with the length. Table~\ref{tab:exp:binomcap} shows that our searches with
binomials attain the bound and never exceed it. Quasi-constacyclic codes where $\lambda$ is not an $m$-th
power reach the same distances as quasi-cyclic ones; they only add larger nominal windows.

\begin{table}[h]
\centering
\begin{tabular}{cccc}
\toprule
$q=p$ & $m=rp^{s}$ & largest $\min(d_X,d_Z)$ found & bound of Prop.~\ref{prop:exp:cap}\\
\midrule
$3$ & $6,\,12,\,24$ \ ($s=1$) & $2$ & $2$\\
$3$ & $18,\,36$ \ ($s=2$) & $3$ & $3$\\
$5$ & $10$ \ ($s=1$) & $3$ & $3$\\
$5$ & $25,\,50$ \ ($s=2$) & $4$ & $4$\\
$7$ & $14,\,21,\,28,\,42$ \ ($s=1$) & $4$ & $4$\\
$7$ & $49$ \ ($s=2$) & $5$ & $5$\\
\bottomrule
\end{tabular}
\caption{Quasi-cyclic and quasi-constacyclic codes of the family ($\rho=0$).}
\label{tab:exp:binomcap}
\end{table}

Notice that, over $\mathbb F_p$, one has $t(x)^p=t(x^p)$, so for a trinomial $t$
the polynomial $f=t^p$ is again a trinomial  and has the repeated
roots required by $h^2\mid f$. Proposition~\ref{prop:exp:cap} still gives $d_X\le d(\langle h\rangle)\le
d(\langle t^{(p-1)/2}\rangle)$, but this is larger than the binomial bound: we computed
$d(\langle t^{2}\rangle)=6$ for $t=x^5+2x^2+4$, $x^5+x^2+4$, $x^5+x^2+3$, $x^5+3x+2$ over $\mathbb F_5$, and
$d(\langle t^{3}\rangle)\ge7$ for all $36$ trinomials $t=x^7+x^i+b$ ($1\le i\le6$, $b\neq0$) over $\mathbb F_7$.
Tables~\ref{tab:exp:q5} and~\ref{tab:exp:q7} list the best codes found at the two lengths in~\cite{DuMaLiu2023},
$N=50$ over $\mathbb F_5$ and $N=98$ over $\mathbb F_7$, both for repeated-root trinomials and for binomials.

\begin{table}[h]
\centering
\small
\begin{tabular}{c|cccc|cc}
\toprule
& \multicolumn{4}{c|}{$f=t^5$, $t$ trinomial ($\rho=1$)} & \multicolumn{2}{c}{$f=x^{25}-1$ ($\rho=0$)}\\
$K$ & $\min(d_X,d_Z)$ & $d_1$ & window & max.\ $(a_l,a_r)$ & $\min(d_X,d_Z)$ & $d_1$\\
\midrule
$10$ & $6$ & $6$ & $4$ & $(2,1)$ & $4$ & $3$\\
$12$ & $6$ & $5$ & $20$ & $(2,2)$ & $4$ & $3$\\
$14$ & $6$ & $5$ & $4$ & $(2,1)$ & $4$ & $2$\\
$16$ & $6$ & $5$ & $4$ & $(2,1)$ & $4$ & $2$\\
$18$--$22$ & $6$ & $3$ & $124$ or $20$ & $(1,1)$ & $4$ & $2$\\
$24$ & $6$ & $3$ & $4$ & $(1,1)$ & $4$ & $1$\\
$26,\,28$ & $6$ & $2$ & -- & none & $3$ & $2$\\
$30,\,32$ & $5$ & $\le2$ & -- & none & $3$ & $\le2$\\
\bottomrule
\end{tabular}
\caption{Index-$2$ codes of length $N=50$ over $\mathbb F_5$ ($m=25$). The column $(a_l,a_r)$ gives the
largest window with $b_1=0$ allowed by~\eqref{eq:exp:conditions}; $(2,1)$ may be replaced by $(1,2)$.
The only QSC of this length in the literature is $[[50,4,4]]_5$~\cite{DuMaLiu2023}. Three independent
runs gave the same values of $\min(d_X,d_Z)$ and $d_1$; the windows listed are the largest found.}
\label{tab:exp:q5}
\end{table}

\begin{table}[h]
\centering
\small
\begin{tabular}{llccccl}
\toprule
family & code & $\min(d_X,d_Z)$ & $d_1$ & window & max.\ $(a_l,a_r)$ & $f$\\
\midrule
\cite{DuMaLiu2023} & $[[98,48,5]]_7$ & $5$ & -- & $<49$ & see Remark~\ref{rem:exp:dml} & $x^{49}-1$\\
\cite{DuMaLiu2023} & $[[98,28,5]]_7$ & $5$ & -- & $<49$ & see Remark~\ref{rem:exp:dml} & $x^{49}-1$\\
\midrule
binomial & $[[98,48,5]]_7$ & $5$ & $1$ & $<7$ & $(3,3)$ & $x^{49}-1$\\
binomial & $[[98,32,5]]_7$ & $5$ & $3$ & $<7$ & $(3,3)$ & $x^{49}-1$\\
binomial & $[[98,14,5]]_7$ & $5$ & $4$ & $<7$ & $(3,3)$ & $x^{49}-1$\\
\midrule
trinomial & $[[98,64,7]]_7$ & $7$ & $2$ & -- & none & $x^{49}+x^7+3$\\
trinomial & $[[98,58,7]]_7$ & $7$ & $2$ & -- & none & $x^{49}+x^7+1$\\
trinomial & $[[98,54,7]]_7$ & $7$ & $3$ & $<16$ & $(1,1)$ & $x^{49}+x^7+1$\\
trinomial & $[[98,50,7]]_7$ & $7$ & $4$ & $<48$ & $(1,1)$ & $x^{49}+x^7+2$\\
trinomial & $[[98,42,7]]_7$ & $7$ & $5$ & $<48$ & $(2,2)$ & $x^{49}+x^7+3$\\
\bottomrule
\end{tabular}
\caption{Index-$2$ codes of length $N=98$ over $\mathbb F_7$ ($m=49$). Here
$x^{49}+x^7+b=(x^7+x+b)^7$. The trinomial rows come from $3$ of the $36$ trinomials, with $10$ pairs
$(g,h)$ each, so better codes may exist.}
\label{tab:exp:q7}
\end{table}

\begin{example}\label{ex:exp:best}
In the three cases the multiplier $k$ is listed in the data files of the repository.
\begin{enumerate}
\item $f=x^{49}+x^7+1=(x^7+x+1)^7$ over $\mathbb F_7$,
$g=x^{6}+3x^{5}+x^{4}+4x^{3}+x^{2}+3x+5$,
$h=x^{16}+x^{15}+x^{14}+x^{12}+6x^{11}+x^{10}+3x^{8}+4x^{7}+2x^{6}+3x^{5}+x^{4}+4x^{3}+3x^{2}+2x+5$ give a $(1,1)$-$[[98+2,54]]_7$ QSC with
$d_X=d_Z=7$, $d_1=3$. It improves on $[[98,48,5]]_7$ of~\cite{DuMaLiu2023} both in dimension and in
distance.
\item $f=x^{49}+x^7+3=(x^7+x+3)^7$,
$g=x^{12}+4x^{11}+3x^{10}+2x^{9}+x^{8}+3x^{7}+5x^{6}+x^{5}+3x^{4}+x^{3}+x^{2}+x+3$,
$h=x^{16}+3x^{15}+3x^{14}+x^{12}+4x^{11}+2x^{9}+6x^{8}+2x^{6}+2x^{5}+x^{4}+x^{3}+5x^{2}+5x+3$ give a
$(2,2)$-$[[98+4,42]]_7$ QSC with $d_X=d_Z=7$, $d_1=5$, $14$ more logical qudits than
$[[98,28,5]]_7$.
\item $f=x^{25}+2x^{10}+4=(x^5+2x^2+4)^5$ over $\mathbb F_5$,
$g=x^{9}+4x^{8}+x^{7}+3x^{6}+2x^{5}+x^{4}+3x^{3}+2x^{2}+4x+1$,
$h=x^{10}+4x^{7}+3x^{5}+4x^{4}+x^{2}+1$ give a $(2,2)$-$[[50+4,12]]_5$ QSC with $d_X=d_Z=6$, $d_1=5$,
whereas every binomial code of the family at this length has $d\le4$.
\end{enumerate}
\end{example}

\begin{table}[h]
\centering
\small
\begin{tabular}{llccc}
\toprule
code & type & $s=0$ & $|s|=1$ & $|s|=2$\\
\midrule
$[[98,48,5]]_7$ & binomial, $d_1=1$ & $b_1=0,\ b\le2,\ p\le2$ & same & same\\
$[[98,32,5]]_7$ & binomial, $d_1=3$ & $b_1\le1,\ b\le2,\ p\le2$ & same & same\\
$[[98,54,7]]_7$ & trinomial, $d_1=3$ & $b_1\le1,\ b\le3,\ p\le3$ & $b_1=0,\ b\le3,\ p\le2$ & --\\
$[[98,42,7]]_7$ & trinomial, $d_1=5$ & $b_1\le2,\ b\le3,\ p\le3$ & $b_1\le1,\ b\le3,\ p\le2$ & $b_1=0,\ b\le3,\ p\le2$\\
\midrule
$[[50,12,4]]_5$ & binomial, $d_1=3$ & $b_1\le1,\ b\le1,\ p\le1$ & same & same\\
$[[50,12,6]]_5$ & trinomial, $d_1=5$ & $b_1\le2,\ b\le2,\ p\le2$ & $b_1\le1,\ b\le2,\ p\le2$ & $b_1=0,\ b\le2,\ p\le1$\\
\bottomrule
\end{tabular}
\caption{Correction capability as a function of the misalignment, from Equation~\eqref{eq:exp:conditions}.
The binomial codes tolerate up to $|s|=3$ ($q=7$) or $|s|=2$ ($q=5$) at no cost.}
\label{tab:exp:capability}
\end{table}

Table~\ref{tab:exp:capability} summarizes the trade-off. For $|s|\le1$, which by
Section~\ref{sec:channel} is the most likely case, $[[98,42,7]]_7$ and $[[50,12,6]]_5$ match or
beat the binomial codes of the same length in every entry; $[[98,54,7]]_7$ corrects more bit and phase
flips than both binomial codes, but for $|s|=1$ it allows no bit flip in the synchronization block,
whereas $[[98,32,5]]_7$ allows one. Over $\mathbb F_7$ the trinomial codes also have larger dimension
than the binomial codes with a smaller $d_1$ ($54$ against $48$, and $42$ against $32$). For $|s|=2$
they lose some capacity in the synchronization block, and for $|s|\ge3$ they do not synchronize.
With the symmetric case of Section~\ref{sec:channel},
$\Prob{|s|\ge2}=2p_s^{2}/(1+p_s)$, so the trinomial codes are preferable for small $p_s$.
Note also that in every code of this family the synchronization block is the weak link for bit
flips: $b_1$ is limited by $d_1$, which is much smaller than $d_X$.

\begin{remark}\label{rem:exp:dml}
In the proof of Theorem 1 in~\cite{DuMaLiu2023}, the physical misalignment of the whole length-$2m$ word
seems to be identified with multiplication by $x^{a}$ in every block. If the two blocks are
interleaved, as we understand is the case there, a one-position shift maps $(c_0,c_1)$ to
$(xc_1,c_0)$, which is not the quasi-cyclic shift, and in small examples we found codes of the
family that are not invariant under it. In that case the argument would directly cover only
misalignments that are multiples of the index. The block synchronization of
Theorem~\ref{theo:main3}, in which only one block is read through its own window, avoids this
question.
\end{remark}

Note that
the codes above are good as QSCs, not as quantum codes in absolute terms. For instance, the best
known stabilizer codes~\cite{Grassl} with the parameters of Table~\ref{tab:exp:q5} are
$[[50,28,8]]_5$ and $[[50,10,14]]_5$; they are additive codes over $\mathbb F_{25}$, not CSS, and
Fujiwara's marker mechanism does not apply directly to them. Extending synchronization to non-CSS
quasi-twisted stabilizer codes is the main open problem suggested by these data.

\section*{Acknowledgements}
This paper grew out of the Master's thesis of the second author, defended in September 2026. We
thank Carlos Munuera, Umberto Mart\'inez-Pe\~nas and Diego Ruano for their insightful comments and
guidance during the defence, which helped sharpen this paper. We carried out the computations on the Bourbaki server of the SINGACOM research group.  E. Mart\'inez-Moro is partially supported by Grant SAGACT-1 PID2022-138906NB-C21 (2023/2027), funded by MCIN/AEI/10.13039/501100011033 (Ministerio de Ciencia e Innovaci\'on, Agencia Estatal de Investigaci\'on) and by FEDER ``Una manera de hacer Europa'', and by the Department of Education of the Junta de Castilla y Le\'on and FEDER Funds (Reference: CLU-2025-1-02-IMUVA).

\section*{Data availability}
The Magma and Python code used in this work and the raw output of all searches are available at
\url{https://github.com/edgarmartinezmoro-web/qsc-magma}. Every code in Section~\ref{sec:experimental}
can be rebuilt from the stored polynomials $f,g,h,k$. The scripts \texttt{magma/verify\_examples.m}
and \texttt{magma/verify\_q7\_one.m} check the parameters of the codes in
Tables~\ref{tab:exp:aly}, \ref{tab:exp:aly5}, \ref{tab:exp:binomcap}, \ref{tab:exp:q5}
and~\ref{tab:exp:q7} and in Example~\ref{ex:exp:best}.

\bibliographystyle{amsplain}
\bibliography{referencias}

\end{document}